\documentclass[11pt]{article}
\usepackage{epstopdf}                                                          
\usepackage[a4paper,hmarginratio=1:1,vmarginratio=2:3,totalwidth=15.2cm,totalheight=22.56cm]{geometry}
\usepackage{bm,epstopdf,epsfig,amsmath,amssymb,amsfonts,colordvi,wrapfig,comment,cancel,verbatim,slashed}
\usepackage{epsfig,bbm}
\usepackage{graphicx,graphics}
\usepackage[font=md,captionskip=8pt]{subfig}
\usepackage[usenames,dvipsnames]{color}
\usepackage[noadjust]{cite}
\usepackage{xcolor} 
\usepackage[utf8]{inputenc}
\usepackage{setspace}

\usepackage[utf8]{inputenc}

\newcommand{\q}{\alpha}

\newcommand{\w}{\omega}

\newcommand{\tGamma}{\tilde\Gamma}

\allowdisplaybreaks

\newcommand{\cD}{{\cal D}}

\newcommand{\cI}{{\cal I}}

\newcommand{\cK}{{\cal K}}

\newcommand{\cO}{{\cal O}}

\newcommand{\ra}{\rightarrow}

\newcommand{\be}{\begin{equation}}
\newcommand{\ee}{\end{equation}}
\newcommand{\bea}{\begin{eqnarray}}
\newcommand{\eea}{\end{eqnarray}}
\newcommand{\Ra}{\Rightarrow}

\newcommand{\baa}{\begin{array}}
                    \newcommand{\eaa}{\end{array}}

 \usepackage{hyperref}  

\long\def\symbolfootnote[#1]#2{\begingroup
\def\thefootnote{\fnsymbol{footnote}}\footnote[#1]{#2}\endgroup}

\begin{document} 
\begin{flushright}
\end{flushright}
\bigskip\medskip
\thispagestyle{empty}
\vspace{2cm}

\begin{center}
\vspace{0.5cm}


\bigskip

{\Large \bf  Weyl conformal  geometry vs  Weyl anomaly}

 \vspace{1.5cm}
 
 {\bf D. M. Ghilencea}
 \symbolfootnote[1]{E-mail: dumitru.ghilencea@cern.ch}
 
\bigskip 

{\small Department of Theoretical Physics, National Institute of Physics
 \smallskip 

 and  Nuclear Engineering (IFIN), Bucharest, 077125 Romania}
\end{center}

\medskip

\begin{abstract}
  \begin{spacing}{0.99}
    \noindent
Weyl conformal geometry is a gauge theory of scale invariance that naturally
brings together the Standard Model (SM) and Einstein gravity. The SM embedding in
this geometry is possible without  new degrees of freedom beyond  SM and Weyl geometry, while
Einstein gravity  is generated  by the broken phase of this symmetry. This follows
a Stueckelberg breaking mechanism in  which the Weyl gauge boson becomes massive
and decouples, as discussed in the past
(\href{https://inspirehep.net/literature/1710395}{arxiv:1812.08613},
\href{https://inspirehep.net/literature/1729735}{1904.06596},
\href{https://inspirehep.net/literature/1861597}{2104.15118}).
However,  Weyl anomaly could break explicitly this gauge symmetry, hence we study it
in Weyl geometry. We first  note that in Weyl geometry {\it  metricity}  can be restored
with respect to a new differential operator  ($\hat \nabla$)
that also enforces a Weyl-covariant formulation. This leads to  a metric-like
Weyl gauge invariant formalism  that enables one to do  quantum calculations directly in Weyl
geometry, rather than use a Riemannian (metric) geometry picture. The result is  the
Weyl-covariance in $d$ dimensions of all  geometric operators ($\hat R$, etc)
{\it and} of their derivatives ($\hat\nabla_\mu\hat R$, etc),  including the
Euler-Gauss-Bonnet term. A natural Weyl-invariant  dimensional regularisation
of quantum corrections exists  and  Weyl gauge symmetry is then maintained
and manifest  at the quantum level, in $d$ dimensions. This is  related to a
non-trivial current of this symmetry,  the divergence of which cancels the
trace  of the energy-momentum tensor. The "usual" Weyl anomaly and Riemannian
geometry are recovered in the (spontaneously) broken phase. The  relation to
holographic Weyl anomaly is discussed.
\end{spacing}
\end{abstract}

\newpage

\section{Motivation}\label{1}
Scale symmetry in its various forms (global, local/gauged) may play a role in physics 
beyond the Standard Model (SM). For example  at high energies e.g. in the early Universe,  
 the SM states are essentially massless and the theory has such symmetry.
Moreover,  the SM with the Higgs mass parameter set to zero is scale invariant  \cite{Bardeen}. 
In this work we consider  a local, {\it gauged} scale symmetry, also known as Weyl gauge symmetry. 
While  it does not have such symmetry, Einstein gravity can actually be a spontaneously broken 
phase of this symmetry in  a more fundamental  theory.  One such theory is the 
original Weyl quadratic gravity  which is a gauge theory of  scale invariance - a symmetry 
inherited  from its underlying Weyl conformal geometry  \cite{Weyl1,Weyl2,Weyl3}.
Using the gauge principle,  Weyl geometry then brings together SM and
Einstein gravity in a gauge theory of scale invariance, as outlined below.

In Weyl  geometry, both the Weyl connection and the spin connection have a Weyl gauge symmetry.
Hence, this geometry is a well-suited framework to study this symmetry. Then in
 theories with this symmetry both  the action and its underlying geometry i.e. its associated
gravity,  naturally share this symmetry. This is important since ultimately geometry "is" physics and
 in curved space-time one cannot really "separate" the action from the geometry\footnote{This 
   is an advantage compared to  Weyl symmetric theories (which have no gauge boson)
   in Riemannian geometry  where  its Levi-Civita connection
 does not have the symmetry of the action.}.
This geometry generates the Weyl quadratic gravity theory  \cite{Weyl1,Weyl2,Weyl3}.

From this theory,  Einstein gravity and a small positive cosmological
constant  are naturally obtained \cite{Ghilen0}
in the  broken phase of this symmetry  after a  Stueckelberg mechanism: the 
Weyl gauge boson ($\omega_\mu$) of scale invariance becomes a  massive  Proca field\footnote{
This result dismissed \cite{Ghilen0}  long-held 
criticisms since Einstein \cite{Weyl1} against Weyl geometry/gravity  as a physical theory
(criticised for its non-metricity $\tilde\nabla_\mu g_{\alpha\beta}\!\not=\!0$),
based on a  wrong implicit assumption of  a {\it massless} 
Weyl gauge field $\w_\mu$. Actually,  $\w_\mu$ is massive \cite{Ghilen0,SMW} and
non-metricity effects are  strongly suppressed by the mass of $\w_\mu$,
$m_w\!\propto\!\! M_{\rm {Planck}}$  \cite{Ghilen0,SMW}; current 
lower bounds on non-metricity scale are low, $m_\w\! \geq\! 1$ TeV \cite{Latorre}
(one cannot define a clock rate in the absence of a mass scale in the symmetric phase,
to claim a second clock effect in this phase).
Further, Weyl geometry is semi-metric i.e.  the {\it ratio} of lengths of two vectors
 (instead of length itself) is  constant under the parallel transport, in agreement with 
an expectation that physics is independent of units of length/time see e.g. \cite{Ghilencea:2022}.
This is  similar to local conformal models where if we use
only light rays to do measurements absolute lengths and time spans cannot be observed \cite{tHooft}.
Finally,  Weyl theory discussed here  is  metric with respect to a new
Weyl-covariant differential operator that we define to implement manifest Weyl
covariance, taking account  of Weyl charges of curvature tensors/scalar, see later.
This preserves the norm of a vector under a {\it Weyl gauge covariant}
parallel transport \cite{Ghilencea:2022} (Appendix B), \cite{Lasenby}.},
 after  "eating" the dilaton  propagated by the $\tilde R^2$ term of Weyl quadratic
gravity \cite{Ghilen0}, and then decouples.
Further, one can show that the  SM (with  higgs mass parameter set to zero) has 
a natural, truly minimal embedding  in Weyl  geometry \cite{SMW,SMW2} with
no new states beyond SM and Weyl geometry. This gives an interesting  {\it UV completion 
of SM and Einstein gravity} in a  gauge theory of scale invariance.
In such case only the Higgs field of  SM acquires a tree-level
coupling to $\omega_\mu$ and may be generated in the early 
Universe by  Weyl boson fusion $\omega_\mu \omega_\mu\!\ra\! h h$ \cite{SMW}.
With $\omega_\mu$ of geometric origin, one can explain the origin of mass by  Weyl 
geometry \cite{Ghilencea:2022}.  Successful inflation is obtained \cite{WI1,WI2,WI3}
giving a gauged version of Starobinsky inflation~\cite{Starobinsky}.

What happens at quantum level?
With scale symmetry a (quantum) gauge symmetry, an immediate question arises. It is well-known
that Weyl symmetry is anomalous \cite{Duff,Duff2,Duff3,Deser1976}, 
so one should address how Weyl anomaly is reconciled with  a  gauged scale symmetry. 
This issue arises partly because quantum corrections (more correctly, their regularisation)
do not respect this classical symmetry. For example, in  dimensional regularisation (DR)  the
analytic continuation  to $d\!=\!4-2\epsilon$ dimensions   breaks explicitly  this symmetry.
Weyl invariant regularisations could address this matter \cite{Englert}, but Weyl
anomaly is more than a regularisation issue: it involves the Euler-Gauss-Bonnet term
which is not Weyl-covariant in $d$ dimensions. Also our Weyl gauge symmetry has a current
- what is its role? these are the right questions to ask, to answer how
Weyl gauge symmetry can avoid  Weyl anomaly.

This brings us to another   motivation for this study. In an  interesting work \cite{Misha2,Misha3}
in a flat spacetime approach  it was shown that the only anomaly-free  Weyl
symmetry in the presence of  dynamical gravity is  global scale invariance.
This result uses an effective action of
the dilaton in a flat space-time limit that arises from general quadratic
curvature actions. So can we avoid this no-go theorem? The reason this
result does not apply here is that  we have a different symmetry, current and
geometry - Weyl geometry (WG) and then:
a)  the dilaton is absent at low scales, having been "eaten" by the Weyl-Proca gauge field to {\it all orders};
b)~there is a  Weyl {\it gauge} symmetry  which has  a non-trivial current, the divergence of which
cancels   the energy-momentum tensor trace $T_\mu^\mu$;
c)  Very important, in  WG the Euler-Gauss-Bonnet operator is Weyl-covariant 
in $d$ dimensions;  e) Finally,  models in WG
have a de Sitter ground state, with a small  cosmological constant
$\Lambda=3 H_0^2>0$ \cite{SMW} preventing an exact flat spacetime limit
approach. We  detail  these results in this work.

Finally, another  motivation for an interest in Weyl anomaly in Weyl geometry comes
from a holographic perspective
of AdS/CFT. In recent \cite{Jia,Ciambelli} it was shown that  Weyl geometry (in $d$ dimensions) is induced 
on the conformal boundary of a $d+1$ asymptotically locally anti-de Sitter (AlAdS) spacetime;
the local Weyl symmetry and Weyl geometry are induced
by diffeomorphism invariance in the bulk spacetime when working in the Weyl-Fefferman-Graham
(WFG) gauge \cite{Ciambelli,Jia}.
Unlike in a Fefferman-Graham (FG) gauge where a bulk Levi-Civita (LC) connection induces
on the conformal boundary also a LC connection (of the boundary metric), in the WFG gauge the bulk LC connection
induces on the boundary the Weyl-covariant geometry. The induced metric and Weyl connection
act as non-dynamical backgrounds of the dual quantum field theory\footnote{As in 
 FG gauge, the boundary metric sources the boundary energy momentum tensor. But  Weyl connection does
not source a current, since it originated from a pure gauge mode of  the bulk metric \cite{Ciambelli,Jia}}. 
Nevertheless, the Weyl  connection makes the geometric quantities on the boundary Weyl-covariant.
As a result,  Weyl anomaly in the 4D boundary in the WFG gauge (while of 
similar form to that in FG gauge)  has now become Weyl-covariant.  We   recover this result, but
we have in addition a dynamical connection and  non-trivial current (as mentioned), relevant for
the anomaly.

These  motivations and the good results so far of Weyl gauge invariance as a 
symmetry and gauge principle beyond SM and Einstein gravity,  justify this  study in  Weyl geometry.

The plan of the paper is as follows: section~\ref{2} reviews Weyl anomaly and 
 regularisations that respect Weyl gauge symmetry. New results are shown in Section~\ref{3}:
 we first note that Weyl geometry can be made {\it metric} with respect to a new differential
 operator ($\hat \nabla$), in which case  geometric operators (curvature tensors/scalar)
 and their {\it derivatives} in $d$ dimensions are  Weyl-covariant
 (this includes the Euler-Gauss-Bonnet operator).
 This enables  a {\it metric-like} geometry   similar to  Riemannian geometry,
 that  preserves Weyl gauge invariance of the action; it allows quantum calculations in Weyl geometry
 that would otherwise require one use  a Riemannian picture, as\,usual.
 With these ideas and symmetry  we show by construction, how
 Weyl gauge symmetry is maintained at the quantum level in Weyl geometry-based models
 and how Weyl anomaly  is generated by the (spontaneously) broken phase.
 Conclusions are in Section~\ref{4} and the Appendix gives additional technical details.

\section{Weyl anomaly and beyond} \label{2}

Let us first review  the Weyl anomaly  \cite{Duff,Duff2,Duff3,Deser1976}.
By {\it Weyl symmetry} we mean the invariance of the action  under a transformation 
of the metric $g_{\mu\nu}$ and, if present,  of its scalar(s) $\phi$ and
fermion(s) $\psi$, as shown below in $d$ dimensions
\medskip
\bea\label{WGi}
g_{\mu\nu}'=\Sigma^q g_{\mu\nu}, \,\,\,
\phi'=\Sigma^{q_\phi} \phi,
\quad
\psi'=\Sigma^{q_\psi}\psi,
\quad
q_\phi=-\frac{d-2}{4}q,
\quad
q_\psi=-\frac{d-1}{4}q,
\,\,\,\Sigma=\Sigma(x)
\eea

\medskip\noindent
Here $q$ is the Weyl charge of the metric. Its normalization is  arbitrary and
conventions for $q$ vary\footnote{
  Note  that in Weyl geometry the mass of Weyl gauge boson
  is proportional to the charge $q$ of $g_{\mu\nu}$ \cite{Ghilen0,SMW}.},
e.g. $q=1$ in \cite{Smolin,Ghilen0} or $q=2$ in \cite{Kugo}; if $q=2$ and $d=4$ 
we have  $q_\phi=-1$,  $q_\psi=-3/2$ i.e. the fields'  dimension in length units.
For simplicity, one can set below $q=2$.

Following \cite{Duff},
consider a  Weyl invariant  action  of some (unspecified) massless matter fields,
in interaction with an external gravitational field and an external spin-1 gauge field. 
Assuming  the absence of a self-interaction of these matter fields, one obtains at one-loop
a gravitational effective action induced by (divergent) quantum contributions from them.
In  dimensional regularisation (DR) this part of\,the action
 has a structure \cite{Duff}\footnote{Our 
conventions are as in \cite{book,Buch}:
$g_{\mu\nu}=(+,-,-,-)$, $g=\vert\det g_{\mu\nu}\vert$,
$R^\mu_{\,\,\,\nu\rho\sigma}=\partial_\rho\Gamma^\mu_{\nu\sigma}+...$,
$R_{\nu\sigma}=R^{\mu}_{\,\,\nu\mu\sigma}$.}
\medskip
\bea\label{Ad}
W_d=\frac{1}{d-4}\int d^d x \sqrt{g}\, A(d),\qquad d=4-2\epsilon,
\eea

\medskip\noindent
where $g=\vert \det g_{\mu\nu}\vert$ and $A(d)$ is a function of the metric and its derivatives. $A(d)$ 
contains higher derivative  operators  such as 
$
R \,\Box^{(d-4)/2} R,$
$R_{\mu\nu}\Box^{(d-4)/2} R^{\mu\nu}$, 
$R_{\mu\nu\rho\sigma}\Box^{(d-4)/2} R^{\mu\nu\rho\sigma}$, 
$F_{\mu\nu} \Box^{(d-4)/2}\, F^{\mu\nu}$,
$C_{\alpha\beta\gamma\delta} \Box^{(d-4)/2} C^{\alpha\beta\gamma\delta}$,
etc, in a Weyl invariant combination, so
the one-loop divergent $W_d$ is Weyl invariant \cite{Buch}.
Here we used $\Box=\nabla^\mu \nabla_\mu$ in Riemannian notation.

The  counterterm action $W_c$  can be written in a basis of independent operators
as below, depends on the DR subtraction scale $\mu$ and has the general structure
 \cite{Duff}
 \medskip
 \bea\label{Wc}
 W_c=-\frac{\mu^{d-4}}{d-4}\int d^d x \sqrt{g}\, \big(
 b K + b' G  + c H\big)
\eea
where $b$, $b'$, $c$ are constants and
\bea\label{not}
 K\equiv C_{\mu\nu\rho\sigma}^2,\qquad
G\equiv R_{\mu\nu\rho\sigma}^2 -4 R_{\mu\nu}^2+R^2\ra E_4, \,\, (d\ra 4),\qquad
H\equiv F_{\mu\nu}^2.
\eea

 \medskip\noindent
 $E_4$ is a total derivative but $G$ is not - this is the Euler-Gauss-Bonnet term;
$C_{\mu\nu\rho\sigma}^2$ is the  Weyl tensor-squared in $d$ dimensions;
 $F_{\mu\nu}$ is a field strength of the  Abelian (external) vector field.  $W_c$ 
is chosen such as  its  pole $1/(d-4)$  cancels that in $W_d$ so the total action is finite.

The trace of the  energy-momentum tensor given by $W_d+W_c$ is
\medskip
\bea
T^\mu_\mu=\frac{-2}{\sqrt{g}}g_{\mu\nu}\frac{\delta (W_d+W_c)}{\delta g_{\mu\nu}}\Big\vert_{d\ra 4}
=\frac{-2}{\sqrt{g}} g_{\mu\nu} \frac{\delta  W_c}{\delta g_{\mu\nu}}\Big\vert_{d\ra 4}
\eea\

 \smallskip
\noindent
where we used that $W_d$  is Weyl invariant.
Further \cite{Duff}
\bea\label{eb}
\frac{2}{\sqrt g} g_{\mu\nu}\frac{\delta}{\delta g_{\mu\nu}}
\int d^d x \sqrt{g} \, K
&=&(d-4) \big( K
+ \frac23 \,\Box R\big),
\\
\frac{2}{\sqrt g} g_{\mu\nu}\frac{\delta}{\delta g_{\mu\nu}}\int d^d x \sqrt{g}\, G&=&(d-4)\, G,\label{eb2}
\\
\frac{2}{\sqrt g} g_{\mu\nu}\frac{\delta}{\delta g_{\mu\nu}}\int d^d x \sqrt{g} \, H &=&(d-4)\, H.\label{eb3}
\eea

\medskip\noindent
For convenience let  us outline here a derivation of (\ref{eb}) \cite{Buch}. 
With notation (\ref{not}) we have
 $\sqrt{g^\prime} \, K (g_{\mu\nu}') \!=\! \sqrt{g} \, K (g_{\mu\nu})\, \Sigma^{q (d-4)/2}$.
Next, for  a functional
$A(g'_{\mu\nu})$  under transformation (\ref{WGi})
\medskip
\bea\label{fun}
\delta A/\delta \ln\Sigma^q=(\delta A/\delta g'_{\mu\nu}) (\delta g'_{\mu\nu}/\delta\ln \Sigma^q)=
(\delta A/\delta g'_{\mu\nu})  g_{\mu\nu} \Sigma^q=g'_{\mu\nu}\delta A/\delta g'_{\mu\nu}.
\eea

\medskip\noindent
For an infinitesimal transformation (\ref{WGi}) 
($\Sigma\!\ra\! 1$, $g'_{\mu\nu}\!\ra\!  g_{\mu\nu}$)
 and with   $\cI(K)\equiv\int d^dx \sqrt{g'} K(g')$ we  find  
$(2/\sqrt{g'}) \,\, g'_{\mu\nu}(\delta/\delta g'_{\mu\nu}) \cI (K)
=(2/\sqrt{g'})\,\, \delta \cI(K)/\delta \ln\Sigma^q\vert_{\Sigma\ra 1}
= (d-4) K$ which was used in (\ref{eb}). This misses the
 $\Box R$ term due to an ambiguity in the local conformal case
 \cite{Duff,Asorey}, but it is easily accounted for, see e.g. \cite{Buch} (section 17.2.2).

 Using (\ref{eb}), (\ref{eb2}), (\ref{eb3}) in the  expression
 for $T^\mu_\mu$ we obtain a {\it finite} correction: 
the pole in  $W_c$  is cancelled  by the factor $(d-4)$  in these equations;
one then takes the limit $d\ra 4$, then
\medskip
\bea\label{anomaly}
T_\mu^\mu=
 b \, [\,C_{\mu\nu\rho\sigma}^2+ (2/3) \Box R\,]+ b' G +c H
 \eea

 \medskip
Since  $T_\mu^\mu\not=0$,  Weyl symmetry is anomalous and a  Weyl invariant 
quantum  theory does not seem possible:
the symmetry was  broken explicitly by the DR scheme in (\ref{Wc})
leading to  the (finite) correction generated in  $T_\mu^\mu$ (the coefficient of  $\Box R$ 
can be altered if an $R^2$ term exists in initial action, whose variation with respect  to $g_{\mu\nu}$ 
will induce $\Box R$;  here no such initial $R^2$ term was assumed, as it would have broken 
initial Weyl symmetry).
The coefficients $b$, $b'$ depend on the matter fields only.
For $N_s$ scalars, $N_f$  fermions and $N_v$ vectors,
 $b=1/(120 k) (N_s+6 N_f +12 N_v)$, $b'=-1/(360 k) (N_s+11 N_f +62 N_v)$,
 $k=(4\pi)^2$.

An attempt to avoid the anomaly  due to the explicit breaking of Weyl symmetry by the DR scheme
and  to realise a quantum conformal gravity (with a spontaneous-only
breaking  of this symmetry) was first made in \cite{Englert}. The authors of  \cite{Englert}
considered (massless) QED corrections to conformal gravity.  It was shown that one could 
avoid  the  anomaly (of type B \cite{Deser}) associated with the Weyl term
$\int d^4 x \sqrt{g} \,C_{\mu\nu\rho\sigma}^2.$
This was possible with an analytical continuation 
different from the DR scheme used in  (\ref{Wc}),
in order to preserve Weyl symmetry  in $d\!=\!4-2\epsilon$. This is done with
the aid of the scalar field (dilaton) $\phi$ with
a Weyl invariant action in $d$ dimensions
\smallskip
\bea
W_\phi=-\int d^d x \sqrt{g} \,\frac{1}{12}\, \Big[\phi^2 R+ \frac{4(d-1)}{d-2} g^{\mu\nu}
\partial_\mu\phi\, \partial_\nu\phi\big].
\eea

\medskip\noindent
 $\phi$ transforms as in (\ref{WGi}), so $\ln \phi\ra \ln\phi +q_\phi\ln\Sigma$
i.e.  $\ln\phi$  has a shift symmetry. When $\phi$ has a (non-zero) constant vev,
Einstein gravity is recovered. The field $\phi$ is used to replace the 
subtraction scale $\mu$  in the counterterm $W_c$ shown below,
to maintain Weyl symmetry \cite{Englert}
\bea\label{weyl}
W_c =-\frac{1}{d-4}\int d^d x \sqrt{g}\, \,\phi^{2(d-4)/(d-2)} \,b\, C_{\mu\nu\rho\sigma}^2.
\eea

\medskip\noindent
Then the simple pole in $W_d$ of (\ref{Ad}) is cancelled as before by  $W_c$, while
$\mu$ is generated spontaneously when  the dilaton acquires a vev, $\mu\! \sim\! \langle\phi\rangle$. 

In this way $W_c$ respects symmetry (\ref{WGi}), so the variation
of $W_c$ is now  $\delta W_c/\delta \ln\Sigma=0$, hence there is no contribution from
 $C_{\mu\nu\rho\sigma}^2$   to (\ref{anomaly})  and  there is no anomaly from (\ref{weyl}):
 essentially, a mixing of $\phi$  with the (external) graviton (see 
 figs. 4 and 6 in \cite{Englert}) leads to  Weyl anomaly cancellation.
 This result  led to various interpretations   \cite{Duff}. However, 
the absence of its associated anomaly  here 
is nothing magic: the theory has an  {\it additional} dynamical degree of freedom (dilaton $\phi$). 
Its decoupling restores the anomaly contribution to $T_\mu^\mu$ \footnote{
The fate of the dilaton  being eaten by $g_{\mu\nu}$  as speculated in
\cite{Englert} is not correct since  it  decouples in the Einstein frame where
 the symmetry is broken. In Weyl geometry the dilaton  is\,eaten by $\omega_\mu$ \cite{Ghilen0}, see later.}.
Hence, Weyl anomaly as we know it merely signals the missing ({\it decoupling}) of an additional
degree of freedom that would otherwise enable the symmetry at  quantum level.

Let us explain  in a different way this absence of the anomaly of the Weyl term
alone in this Weyl-invariant regularisation,  using  non-local form factors ($\ln\Box$) to which it
 is related \cite{Buch,Donoghue}.  In the usual approach  with the DR scale
$\mu$ (instead of a dilaton) as a regulator, the renormalised action $W_r=W_d+W_c$ for the
Weyl term alone  \cite{Buch} is shown below, where 
 $\ln \Box$ ($\ln\mu^2$) arise from $W_d$ ($W_c$), respectively, see eqs.(\ref{Ad}), (\ref{Wc})
\footnote{This follows after
divergence cancellation and after expanding $\Box^{-\epsilon}$ and
$\mu^{-2\epsilon}$ for small $\epsilon$ in eqs.(\ref{Ad}), (\ref{Wc}).}$^,$\footnote{An
analogy to (\ref{1L}) is the conformal anomaly
in QED  where we have $F_{\mu\nu} (1\!+\! \ln\Box/\mu^2) F^{\mu\nu}$ at one-loop.}\cite{Buch,Donoghue}
\medskip
\bea\label{1L}
W_{r}=\frac12\int d^4x \sqrt{g}\,b\,C_{\mu\nu\rho\sigma} \ln (\Box/\mu^2)\, C^{\mu\nu\rho\sigma},
\qquad (b=\textrm{constant}).
\eea

\medskip\noindent
Under (\ref{WGi}), in $d=4$,   $\sqrt{g'}\, C^{'\,2}_{\mu\nu\rho\sigma}\!=\!\sqrt{g} \, C_{\mu\nu\rho\sigma}^2$
but $\Box'=\Sigma^{-q} (\Box+f(\ln\Sigma))$;  here $f(\ln\Sigma)$ depends on the derivatives 
and is neglected for an infinitesimal variation in  (\ref{WGi}), when
 $g_{\mu\nu}\ra g^\prime _{\mu\nu} (\Sigma\ra 1)$, used below.  
Next, for a functional  $A(g'_{\mu\nu})$  we have
eq.(\ref{fun}),  then
\bea T_\mu^\mu = - (2/\sqrt{g'})\,\,  g'_{\mu\nu}\delta W_r/\delta g'_{\mu\nu}
= - (2/\sqrt{g'}) \,\, \delta W_r/\delta \ln\Sigma^q\vert_{\Sigma\ra 1}
= b \, C_{\mu\nu\rho\sigma}^2.
\eea
 This is exactly the anomaly due to $C_{\mu\nu\rho\sigma}^2$ of (\ref{anomaly}),
 re-derived  using   non-local  $\ln\Box$ term \cite{Buch}.

 Alternatively,  a Weyl invariant regularisation eq.(\ref{weyl}) 
generates in $d\!=\!4$ a {\it finite}
\bea
-\frac{b}{2}\int d^4x\, \sqrt{g}\, C_{\mu\nu\rho\sigma}^2\ln\phi^2.
\eea
Then in eq.(\ref{1L}) the factor  $\ln(\Box/\mu^2)$ is replaced by  $\ln (\Box/\phi^2)$
 which is invariant under an infinitesimal transformation  (\ref{WGi}) in $d=4$,
 then so is the new  $W_r$,  hence there is no anomaly (in this symmetric phase).
 This conclusion
 is reached using the $\ln\Box$ form factor, enforcing the similar
 conclusion  in the text after eq.(\ref{weyl}).
 The anomaly re-appears only after spontaneous breaking of Weyl  symmetry
 of  quantum $W_r$: indeed, with $\phi\!=\!\langle\phi\rangle\!+\delta\phi$,
 Taylor expanding  $\ln\phi$  about $\langle\phi\rangle\!=\!\mu$ and after
 neglecting (decoupling) the series of  dilaton fluctuations $\delta\phi$ suppressed by
 $\langle\phi\rangle$, one recovers eq.(\ref{1L}) and the anomaly.

More recently,  the above Weyl-invariant regularisation \cite{Englert} was rediscovered and implemented
in  (global) scale invariant  and conformal theories  in flat space-time \cite{Misha,Misha4} to
obtain quantum theories with this symmetry broken only spontaneously.
 As said,  the subtraction scale is replaced by the dilaton field and all the
 counterterms can then respect the classical symmetry, as shown in detail at
one-loop \cite{Misha,Ghilen1}, two-loop \cite{Ghilen2} and 
three loops \cite{Ghilen3, Monin}. The absence of a scale anomaly in the quantum action
 is then due to the additional presence of a dynamical degree of freedom in the theory (the
 dilaton). When this field acquires a vev and decouples (negligible fluctuations
relative to its vev) the anomaly is recovered in the broken phase only.
Although the scale anomaly vanishes, that does not necessarily mean
that beta functions of the couplings vanish\footnote{
A vanishing  beta function is a sufficient but not a necessary condition for 
quantum scale invariance.}; beta functions  are now defined with respect to 
the rescaling of  the dilaton; with this definition one can check that
Callan-Symanzik equations are respected at 2-loop
\cite{Ghilen2,Ghilen1} together with Ward identities \cite{Tamarit}.

Returning to the symmetry preserving regularisation  \cite{Englert} for 
$C_{\mu\nu\rho\sigma}^2$, the most general action  can contain
additional terms such as the Euler-Gauss-Bonnet term
$G$  in which case the Weyl anomaly (of type A \cite{Deser}) cannot be avoided due to the
explicit breaking of Weyl symmetry in $d$ dimensions: in such case $G$ is not a total derivative
and its action not Weyl invariant - then its anomaly  cannot be removed  by local counterterms
and some special regularisation  (unlike type B, it is $\mu$ independent). This situation
will change in Weyl geometry.

\section{Weyl anomaly in Weyl geometry}\label{3}

Weyl geometry is {\it non-metric} i.e. $\tilde\nabla_\lambda g_{\mu\nu}\not=0$.
Here we first give a brief review of  Weyl geometry and  make the important observation that
this geometry can actually  be treated as a {\it metric} geometry with respect to a new differential
operator ($\hat\nabla$), so $\hat \nabla_\lambda g_{\mu\nu}=0$; $\hat\nabla_\lambda$  preserves
Weyl-covariance when acting on geometric operators  like curvature tensors/scalar (which
are themselves Weyl-covariant in $d$ dimensions), much like in the matter sector of a
gauge theory. These aspects then help us to explain how Weyl gauge symmetry is reconciled
with Weyl anomaly which is recovered in the (spontaneously) broken phase.

\subsection{Weyl geometry with a metric description}\label{3.1}

Weyl geometry\footnote{For more details on Weyl geometry see  Appendix A in \cite{SMW}
and Appendix~\ref{A} of this work.}
 is defined by classes of equivalence
$(g_{\alpha\beta}, \w_\mu$) of the metric ($g_{\alpha\beta}$)
and the Weyl gauge  field ($\w_\mu$),  related by the  Weyl gauge
transformation shown below  in $d=4-2\epsilon$ dimensions, in the absence
(a) and presence (b) of  scalars ($\phi$) and fermions ($\psi$)
\smallskip
\bea\label{WGS}
 (a) &\quad&
 g_{\mu\nu}^\prime=\Sigma^q 
 \,g_{\mu\nu},\qquad
 \w_\mu'=\w_\mu -\frac{1}{\q}\, \partial_\mu\ln\Sigma, 
\qquad
\sqrt{g'}=\Sigma^{q d/2} \sqrt{g},
\nonumber\\[4pt]
(b) &\quad & \phi' = \Sigma^{q_\phi} \phi, 
\quad
\qquad \psi'=\Sigma^{q_\psi}\,\psi,
 \qquad q_\phi=-\frac{q}{4} (d-2),
\qquad  q_\psi=-\frac{q}{4} (d-1).
\eea

\medskip\noindent
This defines the (non-compact) gauged dilatation symmetry or {\it Weyl gauge symmetry}. This extends
 eq.(\ref{WGi}) which is recovered if $\omega_\mu$ is "pure gauge" or zero  everywhere.
Note that here  $\omega_\mu$ is essentially of geometric origin. 
By definition, in Weyl geometry   we have that:
\medskip 
\bea\label{tildenabla}
(\tilde\nabla_\lambda +q \,\alpha \omega_\lambda) g_{\mu\nu}=0,
\qquad \textrm{where}\qquad
\tilde\nabla_\lambda g_{\mu\nu}=\partial_\lambda g_{\mu\nu}-\tilde\Gamma^\rho_{\mu\lambda} g_{\rho\nu}
-\tilde \Gamma^\rho_{\nu\lambda} g_{\rho\mu}.
\eea

\smallskip\noindent
Weyl connection $\tilde\Gamma_{\mu\nu}^\lambda$ is found by standard calculation or
via  $\tilde \Gamma_{\mu\nu}^\lambda=\Gamma_{\mu\nu}^\lambda\big
\vert_{\partial_\mu \ra \partial_\mu+q\,\alpha \omega_\lambda}$ with $\Gamma_{\mu\nu}^\lambda$ 
 the Levi-Civita (LC) connection $\Gamma_{\mu\nu}^\rho=(1/2) g^{\rho\lambda} (\partial_\mu g_{\nu\lambda}
+\partial_\nu g_{\mu\lambda}- \partial_\lambda g_{\mu\nu})$. One finds
\medskip
\bea
\tilde \Gamma_{\mu\nu}^\lambda
=\Gamma_{\mu\nu}^\lambda
+\alpha'\big[\delta_\mu^\lambda \omega_\nu +
\delta_\nu^\lambda \omega_\mu-g_{\mu\nu} \omega^\lambda\big],\qquad \alpha'\equiv \alpha q/2.
\eea

\medskip\noindent
It is easy to check that  Weyl connection ($\tilde \Gamma$) is
invariant under (\ref{WGS}); the same  is true for the Weyl spin connection
\cite{SMW} (Appendix A). This has important consequences, as seen below.

Let us first show the ``standard''  definition of   curvature tensors in Weyl geometry.
The Riemann tensor in Weyl geometry $\tilde R^{\mu}_{\,\,\nu\rho\sigma}$ 
is found as usual from a  commutator acting on a vector field ($v^\lambda$),
 $[\tilde \nabla_\mu, \tilde \nabla_\nu] v^\lambda\!=\!\tilde R^\lambda_{\,\,\,\rho\mu\nu}\, v^\rho$.
This gives the usual  expression, now in terms of~$\tilde \Gamma$
\smallskip
\bea\label{Rie}
\tilde R^\mu_{\,\,\nu\rho\sigma}
=\partial_\rho\tilde\Gamma^\mu_{\nu\sigma}-\partial_\sigma\tGamma_{\nu\rho}^\mu
+\tGamma_{\rho\lambda}^\mu\,\tGamma^\lambda_{\nu\sigma}
-\tGamma_{\sigma\lambda}^\mu\,\tGamma^\lambda_{\nu\rho}.
\eea

\medskip\noindent
with an explicit form in terms of $\w_\mu$ shown in Appendix~\ref{A}.
Since $\tilde\Gamma$ is invariant under  (\ref{WGS}) then
$\tilde R^\mu_{\,\,\,\nu\rho\sigma}$ is invariant, too
and the same is true for the Ricci tensor of Weyl geometry
$\tilde R_{\nu\sigma}\equiv \tilde R^\mu_{\,\,\,\nu\mu\sigma}$. This is different from
Riemannian geometry where the Riemann
and Ricci tensors transform in a complicated way. One finds that  in $d$ dimensions:
\be\label{Ri}
\tilde R_{\mu\nu}=
R_{\mu\nu} 
+\alpha'\, 
\Big[\frac{d}{2} F_{\mu\nu}-(d-2)\nabla_{(\mu} \omega_{\nu)}
 - g_{\mu\nu} \nabla_\lambda\omega^\lambda\Big]
+\alpha^{\prime 2}  (d-2) (\omega_\mu\omega_\nu -g_{\mu\nu} \omega_\lambda\omega^\lambda),
\ee

\medskip\noindent
with $R_{\mu\nu}$ the Ricci tensor in Riemannian geometry, $\nabla$ is 
that of Riemannian geometry (with LC connection) and $\nabla_{(\mu} \omega_{\nu)}=(1/2)
(\nabla_\mu\w_\nu+\nabla_\nu\w_\mu)$.
The  Weyl scalar curvature  is
\medskip
\bea\label{R}
\tilde R=g^{\mu\nu}\tilde R_{\mu\nu}=R-2 (d-1)\, \alpha^\prime \, \nabla_\mu \omega^\mu 
-(d-1) (d-2) \,\alpha^{\prime 2} \omega_\mu \omega^\mu,
\eea

\medskip\noindent
where $R$ is that of Riemannian geometry.
 Note that $\tilde R$ transforms covariantly
under (\ref{WGS}), like the inverse metric $g^{\mu\nu}$ that enters in its definition.
Further, the field strength of $\w_\mu$, regarded as the  length curvature tensor, is
\bea
F_{\mu\nu}=\tilde \nabla_\mu \omega_\nu 
-\tilde \nabla_\nu \omega_\mu
=\partial_\mu\omega_\nu -\partial_\nu \omega_\mu,
\eea
%
where we used that Weyl connection is symmetric
$\tilde \Gamma_{\mu\nu}^\rho=\tilde \Gamma_{\nu\mu}^\rho$ and $\tilde \nabla_\mu\omega_\nu=
\partial_\mu\omega_\nu-\tilde \Gamma_{\nu\mu}^\rho \omega_\rho$.

Finally, the  Weyl tensor of Weyl geometry
 $\tilde C_{\mu\nu\rho\sigma}$ ($\tilde C^\mu_{\,\,\,\nu\mu\sigma}\!=\!0$)
defined by $\tilde R_{\mu\nu\rho\sigma}$,  is
\smallskip
\bea
\tilde C_{\mu\nu\rho\sigma}\!=\!
\tilde R_{\mu\nu\rho\sigma}
+\frac{1}{d-2} \big( g_{\mu\sigma} \tilde R_{\nu\rho}+g_{\nu\rho} \tilde R_{\mu\sigma}
-g_{\mu\rho} \tilde R_{\nu\sigma}-g_{\nu\sigma} \tilde R_{\mu\rho}\big)
+\frac{ \big(g_{\mu\rho}g_{\nu\sigma}-g_{\mu\sigma} g_{\nu\rho}\big)\tilde R}{(d-1)(d-2)}.
\eea

\medskip
To summarise, under (\ref{WGS})  we have:
\bea\label{tildeB}
\tilde R^{\prime\, \mu}_{\,\,\,\,\nu\rho\sigma}=\tilde R^\mu_{\,\,\,\nu\rho\sigma},\quad
\tilde R_{\mu\nu\rho\sigma}'=\Sigma^{q}\,\tilde R_{\mu\nu\rho\sigma},\quad
\tilde R'_{\mu\nu}=\tilde R_{\mu\nu},   \quad
\tilde R'=\Sigma^{- q}\tilde R,\quad
\tilde C_{\mu\nu\rho\sigma}'=\Sigma^q\, \tilde C_{\mu\nu\rho\sigma}.
\eea
and therefore
\bea\label{WC}
\tilde G'\!=\!\Sigma^{-2 q}\,\tilde G,\quad \text{where}\quad
\tilde G\equiv 
\tilde R_{\mu\nu\rho\sigma} \tilde R^{\mu\nu\rho\sigma}  -4 \tilde R_{\mu\nu} \tilde R^{\mu\nu} 
+ \tilde R^2+(2 d^2-7 d+8)\,\alpha^{\prime\, 2} F_{\mu\nu} F^{\mu\nu},
\eea

\medskip\noindent
as detailed in Appendix~\ref{A}, eq.(\ref{tE}).
$\tilde G$ is the Euler-Gauss-Bonnet term of Weyl geometry.

It can be shown that $\tilde G$ does not change the equations
of motion \cite{Bach} and \cite{Tann} (eq.C1).  In the case
$\w_\mu=0$ ($F_{\mu\nu}=0$)  the familiar Riemannian version of the
Euler-Gauss-Bonnet term is recovered,
$\tilde G\ra R_{\mu\nu\rho\sigma}  R^{\mu\nu\rho\sigma}  -4  R_{\mu\nu} R^{\mu\nu}+R^2$.
To conclude, under (\ref{WGS})
\medskip
\bea\label{WGS2}
X^\prime=\Sigma^{-2 q} 
X, \qquad 
 X
 =\tilde R_{\mu\nu\rho\sigma}^2, \, \,\tilde R_{\mu\nu}^2,\,\,\tilde R^2,
 \,\, \tilde C_{\mu\nu\rho\sigma}^2,\,\, \tilde G,
\,\, F_{\mu\nu}^2.
\eea

\medskip
All these terms are thus  Weyl gauge covariant in $d$ dimensions.
This makes it obvious why Weyl geometry is the right framework for implementing Weyl gauge symmetry,
making it  easy to write a Lagrangian invariant under (\ref{WGS}) using these terms
integrated with a $\sqrt{g}\, d^4x$ measure.
This is unlike in Riemannian geometry where they transform in a complicated way.
The reason for this difference  is the invariance of the Weyl connection. %

Despite this advantage,  the above ``standard'' definition of  curvature tensors
is not satisfactory for Weyl geometry as a gauge theory,
because the partial derivative $\partial_\mu$ in $\tilde \nabla_\mu$
when acting on the tensors fields of  geometric  origin is not Weyl covariant,
while when acting on matter fields, Weyl covariantisation is indeed implemented  in the literature,
see e.g.\cite{SMW}. This different treatment prevents a  consistent  Weyl-covariant approach
to all operators, both geometric and matter fields, and of their derivatives.
One consequence is the
presence of  $F_{\mu\nu}^2$  in both $\tilde C^2$ and $\tilde G$ terms above, 
showing that in this basis of operators, $F_{\mu\nu}^2$,  $\tilde C_{\mu\nu\rho\sigma}^2$, 
$\tilde G$  are not independent. Another effect is that the theory is not metric
($\tilde\nabla_\mu g_{\alpha\beta}\!\not=\!0$) making calculations difficult
 and forcing one to go to a Riemannian (metric) picture to do them.

Since $(\tilde\nabla_\lambda+q\alpha \omega_\lambda) g_{\mu\nu}=0$, where
$q$ is the charge of  the metric $g_{\alpha\beta}$, this suggests that for  any given tensor
$T$,  including $g_{\mu\nu}$,  of Weyl charge $q_T$  ($T'=\Sigma^{q_T} T$) one should introduce
a new differential operator
(we suppress the tensor indices)
\bea
\hat\nabla_\lambda T \equiv (\tilde \nabla_\lambda+q_T  \alpha\, \omega_\lambda) \, T
\eea
which transforms covariantly under (\ref{WGS}), using that $\tilde \Gamma$ is invariant:
$\hat\nabla_\mu^\prime T^\prime=\Sigma^{q_T} \hat\nabla_\mu T$.
Regarding $q_T$,  a given tensor of type $T^{(m)}_{(n)}$ has $q_T=(q/2) (n-m)$ e.g. 
$n=2$, $m=0$ for $g_{\mu\nu}$.

Applying this observation at the more fundamental
level of tetrads,  one defines a  more suitable Riemann  tensor (with a "hat")
from which the  length curvature tensor ($F_{\mu\nu}$) effect is removed,
see  Appendix~\ref{A}.
The new Riemannian tensor $\hat R^\tau_{\,\,\,\nu\rho\sigma}$ of Weyl geometry  is (\ref{hattilde})
 \bea\label{nat}
 \hat R^\tau_{\,\,\,\nu\rho\sigma}=\tilde R^\tau_{\,\,\,\nu\rho\sigma}
 -\alpha^\prime \delta_\nu^\tau \,\, \hat F_{\rho\sigma},
 \eea
with  $\hat F_{\mu\nu}= F_{\mu\nu}=\partial_\mu \w_\nu-\partial_\nu \w_\mu$. Hence
\be\label{nat2}
\hat R_{\mu\nu\rho\sigma}=\tilde R_{\mu\nu\rho\sigma}-\alpha^\prime g_{\mu\nu} \hat F_{\rho\sigma},
\qquad
\hat R_{\nu\sigma}=\tilde R_{\nu\sigma}-\alpha^\prime  \hat F_{\nu\sigma},
\qquad
\hat R=\tilde R.
\ee

\medskip
The quantities with a "hat" have the same transformation as in (\ref{tildeB}).
Using  eqs.(\ref{Rie}), (\ref{Ri}), (\ref{R})
one immediately expresses these curvatures in terms of their Riemannian geometry 
counterparts (Appendix~\ref{A}).
One also shows that the Weyl tensor associated to $\hat R_{\mu\nu\rho\sigma}$ 
is equal to that in Riemannian geometry ($C_{\mu\nu\rho\sigma}$), as shown in
eq.(\ref{CCC})
\bea
\hat C_{\mu\nu\rho\sigma}=C_{\mu\nu\rho\sigma}.
\eea
With these we express  $\tilde G$ of (\ref{WC}) in the new "basis"
(note the position of summation indices):
\bea
\hat G= \hat R_{\mu\nu\rho\sigma} \hat R^{\rho\sigma\mu\nu} - 4 \hat R_{\mu\nu} \hat R^{\nu\mu} +\hat R^2.
\eea
 $\hat G$ is a natural generalisation to
 Weyl geometry of the usual Euler-Gauss-Bonnet term and,
 what is important here, it is Weyl-covariant under (\ref{WGS})
as seen using (\ref{tildeB}), (\ref{nat2}).
Notice that now  there is no contribution of $\hat F_{\mu\nu}^2$ to
$\hat C_{\mu\nu\rho\sigma}^2$ or $\hat G$, which are now  independent -
this is welcome for identifying an action without redundant operators.

To conclude, objects with a ``hat'' transform covariantly\,under (\ref{WGS})\,just\,like those 
in\,(\ref{WGS2})
\bea\label{WGS3}
 X^\prime=\Sigma^{-2 q} X, \qquad
 X
 =\hat R_{\mu\nu\rho\sigma}^2, \, \,\hat R_{\mu\nu}^2,\,\,\hat R^2,
 \,\, \hat C_{\mu\nu\rho\sigma}^2,\,\, \hat G,
\,\, \hat F_{\mu\nu}^2,
\eea

\medskip
\noindent
but the  advantage of this more natural  definition of operators  is that  the
differential operator  $\hat\nabla$ acting on these curvature operators is now Weyl gauge
covariant, too. In particular
 \medskip
 \bea\label{wq}
\hat\nabla_\mu' \hat R^\prime=\Sigma^{-q}\hat \nabla_\mu \hat R,\qquad
\hat\nabla^\prime_\mu\hat\nabla^{\prime}_\nu \hat R'=\Sigma^{- q} \,\hat \nabla_\mu\hat\nabla_\nu \hat R, \quad 
\hat\nabla_\alpha' \hat R_{\mu\nu}'=\hat\nabla_\alpha \hat R_{\mu\nu},\,\,\,
\text{etc,}
\eea

\medskip\noindent
which can be seen using that $\tilde\Gamma$ is Weyl invariant; also note that we now have
that
\bea\label{metr}
\hat \nabla_\mu g_{\alpha\beta}=0,
\eea
%
i.e. the theory is {\it metric} ({\it with respect to}  $\hat\nabla$)
in this natural Weyl "basis" (with a "hat").

To conclude, the formulation using geometric operators with a "hat",
largely overlooked  in the literature
(except \cite{Jia,Tann}) in favour of that in (\ref{tildeB}),     is 
important:  it enables a {\it metric-like} formulation giving
at the same time a manifestly Weyl-covariant 
description of {\it geometric} operators {\it and} 
of  their derivatives (acting on  curvature tensors/scalar of the theory), as in any gauge
theory!\footnote{
For a recent update and details on equivalent formulations of Weyl conformal
geometry   see  \cite{CDA}.}.
This is important,  since it enables us to do {\it metric}-like calculations in Weyl geometry
(e.g. quantum corrections,
see next), which would otherwise require one to  go to a Riemannian picture, as usual.
 Relation (\ref{metr})  gives  a close analogy to  metric
Riemannian  geometry (via $\hat\nabla\! \leftrightarrow\! \nabla$)
with the  advantage of Weyl covariance/Weyl gauge symmetry manifest
(in $d$ dimensions),  as in  any gauge theory.
Weyl conformal geometry is thus a {\it covariantised version} of Riemannian
geometry with respect to the gauged dilatations symmetry!
To implement this symmetry,
one can even take Riemannian results and Weyl-covariantise them using
the "hat" notation. This formalism will be used below.

\subsection{Weyl anomaly in Weyl geometry}\label{3.2}

The  most general  Lagrangian of Weyl geometry/gravity in the absence of matter,
 in the original non-metric formulation  (with a tilde), is \cite{Weyl1,Weyl2,Weyl3}
\bea
W_0=\int d^4x \sqrt{g}    \,\Big\{ a_0\tilde R^2+ b_0 \tilde F_{\mu\nu}^2+
c_0 \tilde C_{\mu\nu\rho\sigma}^2 +d_0 \tilde G\Big\}.
\eea
Using the relations of these operators to those in the new basis (with a hat), such as
(\ref{square}), (\ref{tE}), (\ref{CCC}),  and up to a redefinition
of the couplings such as $b_0$,  this action is invariant
\smallskip
\bea\label{WG}
W_0=\int d^4 x \sqrt{g}\,\Big\{ a_0\hat R^2+ b_0 \hat F_{\mu\nu}^2+
c_0 \hat C_{\mu\nu\rho\sigma}^2 +d_0 \hat G\Big\}.
\eea

Each term in  (\ref{WG})  is separately invariant under the Weyl gauge transformation 
eq.(\ref{WGS}) for $d=4$, see eq.(\ref{WGS3}). 
In a quantum theory, even if one of these terms is not included classically
  it will eventually be generated by 
quantum corrections, hence we included all possible terms\footnote{
$C_{\mu\nu\rho\sigma}^2$, generated in all theories of gravity,
 may lead to non-unitarity due to its higher derivatives acting on $g_{\mu\nu}$.
 In Weyl geometry $g_{\mu\nu}$ and $\tilde\Gamma$ are 
independent, in which case this issue seems to be avoided 
\cite{Wheeler}.}
 for a vacuum action and they are all independent 
 in the natural Weyl basis\footnote{Other terms like $\hat\nabla_\mu V^\mu$, $\hat \Box\hat R$, etc
     give a boundary term.}.
Given the symmetry, any higher dimensional operators are not allowed since there
 is no fundamental scale to suppress  them (except non-polynomial terms
 such as $\hat C_{\mu\nu\rho\sigma}^4/\hat R^2$ etc, not considered here).
 For the  couplings we take
 \medskip
\bea\label{coupling}
a_0=\frac{1}{24 \,\xi^2}, \,\,\xi\ll 1; 
\qquad
b_0=\frac{-1}{4},\qquad
c_0=\frac{-1}{\eta^2},\,\,\,\eta<1.
\eea

\medskip
First, if   $\hat F_{\mu\nu}^2$ were absent ($b_0=0$),  we would have  an 
integrable Weyl geometry i.e. locally $\omega_\mu$ is  "pure gauge" or zero, 
and then it  could  be integrated out via its equations of motion; the theory would
become metric in the Riemannian sense $\nabla_\mu g_{\alpha\beta}=0$  \cite{Ghilencea:2022}
(instead of  current $\hat\nabla_\mu g_{\alpha\beta}=0$).
Since the symmetry allows it,  we keep $\hat F_{\mu\nu}^2$ to
have a general Weyl geometry,  with  a dynamical $\omega_\mu$\footnote{
The term $\hat F_{\mu\nu}^2$ also breaks the special conformal symmetry \cite{Kaku}.}
hence we   set  $b_0=-1/4$ for a canonical kinetic term.
Regarding $a_0$ we take  $a_0\propto 1/\xi^2$ where $\xi\ll 1$ is the perturbative coupling of
Weyl quadratic gravity.  We also take  $c_0=-1/\eta^2$, $\eta<1$ but
 make no assumption about $d_0$.

Action (\ref{WG}) gives a Weyl gauge invariant theory, in 
 a metric formulation ($\hat \nabla_\mu g_{\alpha\beta}=0$)
that is spontaneously broken (via Stueckelberg mechanism)
to an Einstein - Proca action for the dilatation gauge field $\w_\mu$ and a small
positive cosmological constant \cite{Ghilen0,SMW}.

Let us then explore the corrections to 
$W_0$  from some massless matter states with an action with symmetry (\ref{WGS}).
One can consider  the  SM  action which can be endowed with
such Weyl gauge symmetry - this is obtained
by a minimal embedding of the SM in Weyl geometry
(with higgs mass  parameter set to zero); 
this is immediate and natural, with {\it no}  additional degrees of freedom beyond SM 
and Weyl geometry \cite{SMW}.  
The SM fermions and gauge bosons actions are simply
those of flat space-time upgraded to curved space-time by 
multiplying them by $\sqrt{g}$  and they are 
invariant under (\ref{WGS}) for $d\!=\!4$ \cite{SMW}.
The Higgs action is easily made Weyl gauge invariant
and for a Higgs singlet state $h$ has the form\footnote{
  We ignore the self-coupling, not relevant here.} \cite{SMW} (eq.(25))
\smallskip
\bea\label{wh}
W_h\!=\!\!\int\! d^4 x\sqrt{g}\,\Big\{\frac12\,\hat \nabla_\mu h \hat\nabla^\mu h
-\frac{1}{12}\xi_h \, h^2 \hat R\Big\},\quad
 \hat\nabla_\mu h=\big[\partial_\mu  +\alpha \,q_h \w_\mu \big]\,h, \quad q_h= \frac{-q}{4} (d-2).
\eea

\medskip\noindent
Note that now  each term is invariant under (\ref{WGS}) for $d=4$.

With these remarks, we can then consider the quantum corrections to vacuum action (\ref{WG}) due
to the   Weyl gauge invariant action of the Higgs or of  SM action,
or of its massless QED part only as done in conformal gravity \cite{Englert} - the discussion below  is
independent of this choice;
$g_{\mu\nu}$ and also $\w_\mu$ that are part of "geometry" are regarded as external  fields.

At the quantum level,  the divergent vacuum action, denoted $W_d $,
which is Weyl gauge invariant,  will  have a structure similar to
that in eq.(\ref{Ad}) but now each individual operator is actually
Weyl-covariant in the new ``hat'' basis. On dimensional and symmetry grounds
$A(d)$  of (\ref{Ad}) will now include  Weyl-covariant operators (see (\ref{wq}))\footnote{
If $f^\prime\!=\!\Sigma^{q_f}f$ under (\ref{WGS}) then
$\widehat \Box^\prime f^\prime\!=\!\Sigma^{q_f-q}\,\widehat\Box f$,
more generally
$\widehat \Box^{\prime\, k} f^\prime\!=\!\Sigma^{q_f-q \,k}\widehat\Box^k f$, $k$ constant,
$\widehat\Box\!\equiv\!\hat\nabla_\mu\hat\nabla^\mu$
}
\medskip
\bea\label{nablas}
\hat R\, ( \hat \nabla_\alpha\hat \nabla^\alpha)^{(d-4)/2} \hat R,
\qquad
\hat R_{\mu\nu}\, ( \hat \nabla_\alpha\hat \nabla^\alpha)^{(d-4)/2} \hat R^{\mu\nu},
\quad
\hat R_{\mu\nu\rho\sigma}\, ( \hat \nabla_\alpha\hat \nabla^\alpha)^{(d-4)/2} \hat R^{\mu\nu\rho\sigma},\,\,\,
\textrm{etc.}
\eea 

\medskip\noindent
When integrated over  $d$ dimensions, with measure $\sqrt{g}$, each of these operators give a Weyl
gauge invariant action\footnote{These terms
  will also generate finite corrections $\ln (\hat \nabla_\mu\hat \nabla^\mu) \hat R^2$
  when expanding for small $\epsilon$, etc.}.
The associated simple poles $1/(d-4)$  in $W_d$ can be cancelled by a counterterm $W_c$ that is 
Weyl gauge invariant in $d$ dimensions and has a general structure similar to (\ref{weyl})
\medskip
\bea\label{wc}
W_c=- \frac{1}{d-4}\int d^d x \sqrt{g}\,\Big\{a_1\hat R^2+ b_1 \hat F_{\mu\nu}^2+
c_1 \hat C_{\mu\nu\rho\sigma}^2 +d_1 \hat G\Big\}\, \phi^{2 (d-4)/(d-2)},
\eea

\medskip\noindent
where the one-loop coefficients $a_1, b_1, c_1, d_1$ are fixed (beta functions)
by the one-loop divergences and depend  on the matter field content considered,
see e.g. \cite{Donoghue}.

In the light of the previous discussion for the Weyl term eq.(\ref{weyl}), we
 implemented a Weyl-invariant regularisation of $W_c$:
 we replaced the usual DR subtraction scale  $\mu^{-2\epsilon}$ of  (\ref{Wc}) 
by 
$\phi^{2 (d-4)/(d-2)}$ where  $\phi$ is the dilaton field. This is 
the field that linearises the quadratic term $\tilde R^2$ in the action (as
shown later, Section~\ref{3.4})\footnote{For more details see Section~2.5
  and eqs.(29), (32) in \cite{SMW}. See also \cite{Ghilen0}.}.
A side-remark is in order here:
if the higgs field contributes to the vacuum action,
the ``true'' dilaton is actually the radial combination of $\phi$ and the higgs, 
while the "angular" combination of these fields
becomes  the physical (neutral) Higgs field at low scales \cite{SMW}.
To a first approximation we  neglect the higgs field contribution, in which
case the factor in (\ref{wc}) is $\phi$, justifying our notation there.
Therefore,  we did not add   "by hand" any extra field:
the dilaton field  is itself part of the
spectrum and  has a  geometric origin in $\hat R^2$ (as mentioned).

Similar to the previous section where  Weyl-invariance was restored for 
 $\hat C_{\mu\nu\rho\sigma}^2\sqrt{g}$   in $d$ dimensions (and avoided the anomaly),
each term  in $W_c$ is here separately Weyl gauge invariant in $d$ dimensions.  This can
be verified with eqs.(\ref{WGS}), (\ref{WGS3}). 
This is true in particular for  $\hat G \, \phi^{2 (d-4)/(d-2)} \sqrt{g}$.
While in Riemannian geometry (topological) $G$ is a total derivative  in $d=4$,
in Weyl geometry $\hat G$  is Weyl covariant  in 
  $d$ dimensions and its contribution to the action is Weyl gauge invariant
 for this analytical continuation.
Had $\hat G$ not been Weyl covariant  then its contribution in (\ref{wc}) could
not have been made invariant  by   $\phi^{(..)}$ regulator.
This shows the important role played by  Weyl geometry.

There is a more natural analytical continuation of geometric origin for $W_c$,
due to Weyl covariance, that does not use the dilaton as regulator; one
replaces it in (\ref{wc}) by
 \smallskip
 \bea \label{reg}
\phi^{2 (d-4)/(d-2)} \ra \vert\hat R\vert^{(d-4)/2}.
\eea

\medskip\noindent
This only apparently leads to a new regularisation or subtraction scale, since  actually 
the dilaton   has an equation of motion $\phi^2=\vert\hat R\vert$, as shown in
Section~\ref{3.4} (see also recent \cite{DBI,review}).

Action (\ref{WG}),  (\ref{wc}) is now Weyl  gauge  invariant
in $d$ dimensions. %
As shown in Appendix~\ref{B} this has the consequence that 
the energy-momentum  tensor is now cancelled by the
 divergence of the Weyl  current,  so the Ward identity is now:
\medskip
\bea\label{traceT}
T_\mu^\mu-\frac{1}{\alpha'} \nabla_\mu J^\mu=0,  \qquad \alpha'=\alpha q/2,
\eea
The current is  conserved onshell 
\bea
 J^\mu+\nabla_\sigma F^{\sigma\mu} \quad \Ra \quad \nabla_\mu J^\mu=0.
\eea
Here $\nabla_\mu$ is that of Riemannian geometry (with Levi-Civita connection $\Gamma$).
This dilaton current is trivial (vanishes) if $\w_\mu$ is ``pure gauge''
($\w_\mu =(1/\alpha) \partial_\mu\ln\phi^2$) or zero
everywhere  \cite{Ghilencea:2022}\footnote{
  The vanishing\,of $J_\mu$   and thus of its charge means that, 
  ultimately, this case is not really physical \cite{J1,J2}}.
The current has a vanishing divergence if we use the equation of motion of $\w_\mu$.
The form of the current is shown in eq.(\ref{jjj})
$J_\mu=\kappa (\partial_\mu -\alpha\, q\, \w_\mu)\phi^2$
where $\kappa=-\alpha/(4\xi^2)$, where $\phi$ is the dilaton field, used in Appendix~\ref{B} to
linearise the $\hat R^2$ term in the action.

In the notation of Weyl geometry we have
\bea
T_\mu^\mu -\frac{1}{\alpha'}\hat\nabla_\mu J^\mu=0,
\qquad
J_\mu=\kappa \hat \nabla_\mu\phi^2
=\kappa \,(\partial_\mu -\alpha q\, \w_\mu) \phi^2.
\eea
 This current  generalises that present
 in the global scale invariant case $\cK_\mu\propto\partial_\mu\phi^2$ \cite{F1,F2,F3,F4,Bellido}.

The vacuum part of the renormalised gravitational action is $W_r=W_0+W_d+W_c$
and has the form below, using (\ref{WG}), (\ref{nablas}), (\ref{wc}):
\medskip
\bea\label{Wr}
W_r\!\!\!\!\! &= &\!\!\!\!\!\!
\int\!\! d^4 x \sqrt{g}\,\, \Big\{\hat R \,
\Big[a_0+\!\frac{a_1}{2} \ln\frac{\widehat\Box}{\phi^2}\Big] \hat R
+ \! \hat F_{\mu\nu}\, \Big[b_0+\!\frac{b_1}{2} \ln\frac{\widehat\Box}{\phi^2}\Big]\,\hat F^{\mu\nu}
\!\!+  \hat C_{\mu\nu\rho\sigma} \, \Big[c_0+\frac{c_1}{2}
\ln\frac{\widehat\Box}{\phi^2}\Big] \,\hat C^{\mu\nu\rho\sigma}
\nonumber\\[5pt]
&& \qquad\qquad\qquad\qquad
+\,\hat R_{\mu\nu\rho\sigma}\, Z\, \hat R^{\rho\sigma\mu\nu}
-4 \hat R_{\mu\nu}\,Z\,\hat R^{\nu\mu} + \hat R\, Z\,\hat R
\Big\},  
\eea
with a notation\footnote{
Under the one-loop log terms in (\ref{Wr}), (\ref{ZZ}), one can
actually replace $\phi^2\ra \vert\hat R\vert$ by using the classical equation
of motion ($\hat R=-\phi^2$), since the difference of doing so is beyond one-loop.}
\bea\label{ZZ}
Z=d_0+\frac{d_1}{2} \ln\frac{\widehat \Box}{\phi^2}
\eea
%
$W_r$  includes UV non-local terms $\ln\widehat\Box$ where 
$\widehat{\Box}=\hat\nabla_\mu \hat\nabla^\mu.$
All  terms in (\ref{Wr}) remain Weyl invariant,  given that $W_0$, $W_d$
and $W_c$ are invariant. The second line in (\ref{Wr}) is due to the Euler-Gauss-Bonnet term.
If we use equivalent (\ref{reg}), replace $\phi^2\!\ra\!\vert \hat R\vert$ under log terms.

Let us detail  the origin of the log terms; 
 terms in the divergent $W_d$   involved $\widehat\Box^{-\epsilon}$ such as 
 $-1/(2\epsilon)\int d^d x \sqrt{g}\, \hat R\,\widehat\Box^{-\epsilon}\hat R$
 which are invariant under (\ref{WGS}), see (\ref{nablas}).
If one expands it in powers of $\epsilon$ and retains only  the leading 
$-1/(2\epsilon)+(1/2)\ln\widehat \Box+\cO(\epsilon)$, the symmetry is violated at $\cO(\epsilon^0)$ by 
this truncation of the expansion, since $\ln\widehat\Box$ transforms.
The same applies to $W_c$ when expanded in $\epsilon$
 that involves $1/(2\epsilon)-(1/2) \ln\phi^2+\cO(\epsilon)$ terms. 
 The two (finite) log terms combine into an invariant $\ln(\widehat\Box/\phi^2)$, thus keeping
 the Weyl invariance of $W_r$ in $d=4$ ($\widehat\Box$ and $\phi^2$ have  the same
Weyl  charge, equal to $-q$ in $d=4$).
 
 Let us note that the form of $W_r$  can also be ``guessed''
 using only symmetry arguments,  Weyl-covariance of each operator 
 and metricity, by  writing an ``upgraded'' Weyl-invariant version of  the usual
 result in DR such as eq.(395) in \cite{Donoghue} (see also \cite{Donoghue2,Buch}).
Note however, that there is an additional term
beyond $\hat C_{\mu\nu\rho\sigma}^2$,   $\hat F_{\mu\nu}^2$  and $\hat G$:
this is the $\hat R^2$ term that plays a crucial role in the symmetry breaking, 
as we discuss  in Section~\ref{3.4}. 

One can replace the Weyl  curvature terms in (\ref{Wr}) 
in terms of their Riemannian expressions, but then the result is not very illuminating.
The presence of  $\ln\phi^2$   as a coefficient of the various terms, including
the   kinetic term  $\hat F_{\mu\nu}^2$, does not allow a flat space time limit\footnote{
  We have that $\langle\phi^2\rangle\sim \Lambda$, where $\Lambda$ is
  the cosmological constant, see next section.};
this is expected from the spontaneous breaking of the Weyl symmetry which assumes
a non-zero vev of the dilaton  and this leads to a 
non-vanishing value of $\vert \hat  R\vert=\phi^2$.

To conclude, we showed that Weyl-covariance of individual operators in the action
(such as $\hat R^2$, $\hat G$, etc) and of their derivatives in the ``hat''
basis (e.g. $\widehat\Box \hat R$, etc),  together with  a Weyl gauge invariant
regularisation  enabled by  $\hat R$, ensure  that  Weyl gauge
symmetry is manifestly present at the quantum level (not broken explicitly by the anomaly).

\subsection{Relation to holographic Weyl anomaly and Riemannian limit}\label{3.3}
Let us consider here using the ``standard'' regularisation (with a scale $\mu$)
which is the same as having a constant vev for $\phi$  in $W_c$ and
$W_r$. Then  in the last line of eq.(\ref{Wr})
$\ln\langle\phi\rangle\sim\ln\mu$=constant
 simply cancels out; further,  under (\ref{WGS}),
 $\ln\widehat\Box\ra \ln\widehat \Box+\ln\Sigma^{-q}$;
then  the last line in (\ref{Wr})
generates an anomaly
 \bea
-\frac{d_1}{2}\int d^4x \sqrt{g} \,\hat G\ln\Sigma^q.
\eea
 This is  the holographic 
 anomaly from the Euler-Gauss-Bonnet term as found in \cite{Jia,Ciambelli}
in WFG gauge, where Weyl geometry is generated on the boundary.
A good consistency check of our result is  that, in agreement with  the holographic picture
 \cite{Jia,Ciambelli}, the term $\hat G$ is now Weyl-covariant.  
This is due to Weyl geometry and differs from the original anomaly 
in Riemannian space-time.   The main difference between 
\cite{Jia,Ciambelli} and this work  is that here $\w_\mu$ is
 dynamical in which case there  is a non-trivial current \cite{Ghilen0} as seen in (\ref{traceT}),
also mentioned in \cite{Ciambelli}. The presence of this current 
is crucial in Weyl anomaly absence in the symmetric phase since
its divergence cancels  the trace $T_\mu^\mu$.
It would be interesting to have the holographic picture for
a dynamical $\w_\mu$  to compare to our result.

The Riemannian limit  of eq.(\ref{Wr})  can  formally be recovered for  $\w_\mu=0$:
then $\tilde \Gamma_{\mu\nu}^\alpha\ra \Gamma_{\mu\nu}^\alpha$ i.e. Weyl connection becomes Levi-Civita,
then $\hat R\ra R$, $\hat R_{\mu\nu}\ra R_{\mu\nu}$, 
$\widehat \Box\ra \Box$.    If one also formally replaces $\ln\phi\ra\ln\mu$
(corresponding to explicitly broken Weyl gauge symmetry by the DR scheme)
then  $\ln\mu$ cancels out in the Euler-Gauss-Bonnet term of  the  second line 
of (\ref{Wr}) and this  is  restoring the usual (type A) Weyl anomaly, see \cite{Donoghue,Donoghue2}. 
The same applies  for the Weyl-tensor-squared  part of the anomaly (of type B). 
This is exactly  the situation found in the  broken phase of Weyl gauge symmetry 
where $\w_\mu$ becomes massive and thus decouples
(as we clarify shortly); then below the (large) mass of $\w_\mu$  formally $\w_\mu= 0$,
Weyl connection (geometry) becomes Levi-Civita (Riemannian), respectively,
Einstein gravity is recovered \cite{Ghilen0,SMW}, and the ``usual'' 
Weyl anomaly is generated from (\ref{Wr}) in this broken phase.

\subsection{Stueckelberg breaking of the symmetry}\label{3.4}

The results above rely on the spontaneous breaking of Weyl gauge symmetry and
 in particular on the role of the dilaton field. This was
discussed extensively  \cite{Ghilen0,SMW,Ghilencea:2022} but we review
it here  for a self-contained analysis.
The breaking is closely related to  the $\hat R^2$ term
in the action. First let us ignore the presence of higgs/matter action,
eq.(\ref{wh}). Then, at the tree-level one 
 linearises the Weyl-covariant term $\hat R^2$ via a replacement 
$\hat R^2\ra -2 \phi^2 \hat R - \phi^4$
in the action,  where $\phi$ is a scalar field. One obtains in this way an equivalent form of the action.
Here we display only the relevant terms involving $\hat R^2$ and $\hat F_{\mu\nu}^2$ (we ignore
quantum corrections), to obtain an equivalent action
\medskip
\bea\label{aux}
W_r = 
\int d^4 x \sqrt{g}\,\, \Big\{
 a_0 \hat R^2
+ b_0 \hat F_{\mu\nu}^2
+  ... \Big\} 
=
\int d^4 x \sqrt{g}\,\, \Big\{ a_0 \,(-2  \phi^2 \hat R -\phi^4)
+ b_0 \,\hat F_{\mu\nu}^2
+...\Big\}
\eea

\medskip\noindent
The equation of motion for $\phi$ has a solution $\phi^2=-\hat R$ ($\hat R<0$)\footnote{
In a Friedmann-Robertson-Walker universe: $\hat R=-12 H_0^2\!<\!0$
and $\Lambda=3 H_0^2$, $H_0$=Hubble constant \cite{SMW,SMW2}.} which when replaced
back in $W_r$ recovers the initial action.   This  assumes a  non-vanishing vev of $\phi$,
something  already used in the regularisation.
Since $\ln\phi$ transforms with a shift under (\ref{WGS}), it plays the role of the dilaton, as
anticipated.
Next, using  (\ref{Rs}), one can express $\hat R$ in  a Riemannian notation;  the action becomes
\cite{Ghilen0} (see also Section 2.1 in \cite{SMW})
\be\label{z1}
W_r\!=\!\int\!\!
 d^4 x \sqrt{g}\, \Big\{\! - 12 a_0 \Big[\frac16\phi^2 R +(\partial_\mu\phi)^2\Big]
-a_0\phi^4 
+3 \alpha^2 q^2 \phi^2 a_0 \Big[\omega_\mu
-\frac{1}{\alpha\,q}\partial_\mu\ln\phi^2\Big]^2\!
-\frac14\,F_{\mu\nu}^2+...\Big\}
\ee

\medskip\noindent
When $\phi$ acquires a constant vev, $W_r$ becomes\footnote{
Formally, one applies on (\ref{aux}) a special Weyl transformation (\ref{WGS}) in $d\!=\!4$ with 
$\Sigma^q\!\equiv\! \phi^2/\langle\phi\rangle^2$ so $\phi^\prime\!=\!\langle\phi\rangle$.}
\bea
W_r=\int d^4 x \sqrt{g}\, \Big\{- \frac12 M_p^2\,R 
+\frac12 m_\omega^2 \omega_\mu \omega^\mu 
-\Lambda \,M_p^2- \frac14\,F_{\mu\nu}^2
+\cdots\Big\},
\eea
and  (using (\ref{coupling}))
\bea
M_p^2=4 a_0 \langle\phi\rangle^2=\frac{\langle\phi\rangle^2}{6\,\xi^2}, 
\qquad
m_\omega^2=\frac32\alpha^2 \,q^2\,M_p^2, \qquad \Lambda=\frac14\,\langle\phi\rangle^2.
\eea 

Hence we obtained the Einstein-Proca action for the Weyl gauge field which
became massive via Stueckelberg mechanism: $\w_\mu$ has absorbed the $\ln\phi$  field
which is the would-be-Goldstone field (dilaton) of gauged dilatations (\ref{WGS})~\cite{Ghilen0}
\footnote{The field $\phi$  is dynamical \cite{Ghilen0,SMW} as seen from the conserved current
  equation, $\nabla_\mu J^\mu=0$ with eq.(\ref{jjj}).}.
Note that the cosmological constant is much
 smaller than  Planck scale  because gravity is weak ($\xi\ll 1$).\footnote{The
relation $\Lambda/M_p^2\propto\xi^2$ hints at a IR-UV connection corresponding
to these two very different scales.}

After the  massive Weyl gauge field decouples (together with the dilaton), the usual
Weyl anomaly emerges  in the broken phase of the quantum theory, as discussed
(Section~\ref{3.3}).
Note that  the equation of motion of $\phi$,
$\phi^2=-\hat R$, gives after the symmetry breaking  $R=-4\Lambda\not=0$ \cite{SMW,SMW2}.
Then,  $\ln\phi^2$ in (\ref{Wr}) or (\ref{reg}) generates  $\ln\hat R$ terms
which prevent one from taking an exactly flat metric. This is expected since we implicitly
assumed $\langle\phi^2\rangle\not=0$ when we linearised the $\hat R^2$ term.

 Finally, if the higgs is included, this discussion remains valid  with the only change
 that now  $ \phi^2\hat R$ of (\ref{z1})  combines with $h^2 \,\hat R$ term of
 eq.(\ref{wh}) to generate $M_p^2 R$.
 Hence, the  dilaton in the absence of higgs is now replaced by the radial direction in field space
  $\phi^2\ra \phi^2 + \xi^2\xi_h\,h^2$ ($\xi\ll 1$). It is then this combination
  that is used as regulator in (\ref{wc}) as already explained, or  one is using directly
  eq.(\ref{reg}).

\section{Conclusions}\label{4}

We studied at the quantum level the  gauged scale symmetry (also called Weyl gauge symmetry)
that is  built  in Weyl conformal geometry and
 discussed Weyl anomaly in this geometry.

One motivation for this study was that Weyl geometry is interesting since it
naturally brings together the SM and Einstein gravity in a gauge theory:
 as shown in the past,  the  SM (with vanishing higgs mass parameter) and Einstein gravity admit
 a truly minimal embedding in  Weyl geometry {\it without} any new
degrees of freedom beyond the SM and this geometry.  
This leads to a UV completion in a  fundamental, gauge theory of scale invariance 
that recovers both Einstein gravity and  SM in the spontaneously broken phase.
This phase follows a Stueckelberg mechanism in which the Weyl gauge boson
$\w_\mu$ becomes massive after ``eating'' the dilaton $\ln\phi^2$ propagated
by $\hat R^2$ term in the action. Another motivation of this study was that,
as a (quantum) gauge theory, consistency requires it be anomaly free.
A third motivation was to understand the connection with the holographic Weyl anomaly.

Our result was constructed  on the important observation
that there exists a natural Weyl "basis" for the {\it geometric} operators (curvature
tensors and scalar), little used in the literature, that  has an important advantage:
in this ``basis'' one can restore metricity ($\hat\nabla_\mu g_{\alpha\beta}\!=\!0$) with respect to a new, 
{\it Weyl gauge covariant} differential operator ($\hat \nabla$) acting on these operators.
This  gives:  {\bf 1)}  a {\it metric}-like geometry formalism that
enables one to do quantum calculations directly in Weyl geometry without going to a Riemannian
picture as usually done,
and: {\bf  2)}~all individual (geometric)  operators {\it and} their derivatives are now Weyl
gauge covariant (similar to matter fields operators), as in ordinary gauge theories.
Weyl covariance of the operators is important also because it allows a Weyl
invariant regularisation  of geometric origin (with DR scale $\mu$ replaced by scalar curvature $\hat R$).
Weyl gauge symmetry is  then maintained and manifest at quantum level,
 anomaly-free.

This result is possible because in   Weyl geometry the vacuum action in the natural Weyl basis
then contains  only operators (and counterterms, etc) that are individually  Weyl gauge covariant
in $d$ dimensions and this includes the Euler-Gauss-Bonnet term, and their  derivatives. 
From a symmetry viewpoint, this  brings on  equal footing the $\hat G$ 
 and $\hat C_{\mu\nu\rho\sigma}^2$ terms.
After the Stueckelberg symmetry breaking mechanism, $\w_\mu$ decouples and Riemannian geometry
is recovered together with ``usual'' Weyl anomaly, in the  broken phase.

Our result remains  consistent with that of a Weyl anomaly derived from the 
holographic perspective of AdS/CFT in the WFG  gauge, where Weyl geometry
is generated on the conformal boundary but without dynamical $\w_\mu$. 
The Euler-Gauss-Bonnet term $\hat G$  is Weyl-covariant (in $d$ dimensions),
in agreement to our result. Our case however, having a  dynamical $\w_\mu$, has in addition a
non-trivial current as well as a Weyl-invariant regularisation, both relevant for anomaly absence.
Further study is needed of  the role of this current from the holographic view
and of the   Ward identities in Weyl geometry.
The results so far suggest that Weyl conformal geometry can be the right framework for
a fundamental gauge theory and symmetry beyond both the SM and Einstein gravity.

\bigskip
\bigskip

\section*{Appendix}

\def\theequation{A-\arabic{equation}}
\def\thesubsection{A}
\setcounter{equation}{0}
\def\thefigure{A-\arabic{figure}}
\def\thelabel{A}

\subsection{Weyl geometry in d dimensions}\label{A}
 
We  derive here some formulae used in the text.  For 
additional information  see also Appendix~A in \cite{SMW}.
For convenience we re-write here the symmetry transformation in
(\ref{WGS}):
\bea
 (a) &\quad&
 g_{\mu\nu}^\prime=\Sigma^q 
 \,g_{\mu\nu},\qquad
 \w_\mu'=\w_\mu -\frac{1}{\q}\, \partial_\mu\ln\Sigma, 
\qquad
\sqrt{g'}=\Sigma^{q d/2} \sqrt{g},
\nonumber\\[5pt]
(b) &\quad & \phi' = \Sigma^{q_\phi} \phi, 
\quad
\qquad \psi'=\Sigma^{q_\psi}\,\psi,
 \qquad q_\phi=-\frac{q}{4} (d-2),
\qquad  q_\psi=-\frac{q}{4} (d-1).
\eea
Using $\tilde\nabla$ defined in eq.(\ref{tildenabla}),
the Weyl connection  $\tilde\Gamma_{\mu\nu}^\lambda$ is found from
Weyl geometry equation:
 \be
\tilde \nabla_\mu g_{\alpha\beta} =- \alpha \,q \,\w_\mu \, g_{\alpha\beta}.
\ee
With Levi-Civita connection 
 $\Gamma_{\mu\nu}^\lambda=(1/2)\, g^{\lambda\rho}\, (\partial_\mu g_{\nu\rho}
+\partial_\nu g_{\mu\rho}-\partial_\rho g_{\mu\nu})$, then
\bea
 \tilde \Gamma_{\mu\nu}^\lambda&=&
\Gamma_{\mu\nu}^\lambda\big\vert_{\partial_\mu\ra \partial_\mu+\alpha q \w_\mu}
\\[4pt]
&=&
\frac12 \,g^{\lambda\alpha}\,(\cD_\mu g_{\alpha\nu}+\cD_\nu g_{\alpha\mu}-\cD_\alpha g_{\mu\nu}),
\quad \cD_\mu g_{\alpha\beta}\equiv(\partial_\mu+\alpha q \,\w_\mu) g_{\alpha\beta}
\nonumber\\[4pt]
&=&
\Gamma_{\mu\nu}^\lambda
+\alpha'\Big[\delta_\mu^\lambda \omega_\nu +
\delta_\nu^\lambda \omega_\mu-g_{\mu\nu} \omega^\lambda\Big],
\quad \alpha'\equiv \frac{\alpha q}{2}.
\eea

\medskip\noindent
which is symmetric:  $\tilde\Gamma_{\mu\nu}^\lambda=\tilde\Gamma_{\nu\mu}^\lambda$.
The standard approach to find the
Riemann tensor in Weyl geometry, $\tilde R^{\mu}_{\,\,\nu\rho\sigma}$,
is to use a  commutator acting on a vector field ($v^\lambda$),
 $[\tilde \nabla_\mu, \tilde \nabla_\nu] v^\lambda=\tilde R^\lambda_{\,\,\,\rho\mu\nu}\, v^\rho$ which gives
\bea\label{rrr}
\tilde R^\lambda_{\,\,\mu\nu\sigma}
&=&
\partial_\nu\tGamma^\lambda_{\mu\sigma}
-\partial_\sigma\tGamma_{\mu\nu}^\lambda
+\tGamma_{\nu\rho}^\lambda\,\tGamma^\rho_{\mu\sigma}
-\tGamma_{\sigma\rho}^\lambda\,\tGamma^\rho_{\mu\nu}.
\\[5pt]
&=&
R^\lambda_{\,\,\,\mu\nu\sigma}
+\alpha^\prime\Big\{
\delta_\sigma^\lambda \nabla_\nu \w_\mu
-\delta_\nu^\lambda\nabla_\sigma\w_\mu
-g_{\mu\sigma}\nabla_\nu \omega^\lambda
+ g_{\mu\nu} \nabla_\sigma\omega^\lambda
+\delta_\mu^\lambda \,F_{\nu\sigma}\Big\}\nonumber
\\[5pt]
&+& \alpha^{\prime 2}\Big\{
\omega^2 (\delta_\sigma^\lambda\,g_{\mu\nu} 
-\delta_\nu^\lambda \,g_{\mu\sigma})
+\w^\lambda \,(\w_\nu g_{\sigma\mu}-\w_\sigma g_{\mu\nu})
+ \w_\mu \,(\w_\sigma \delta_\nu^\lambda -\w_\nu\,\delta_\sigma^\lambda)\Big\}\label{A6}
\eea
where $R^\lambda_{\,\,\,\mu\nu\sigma}$ is the Riemannian  geometry counterpart,
$\nabla_\mu$ acts with Levi-Civita connection, $\nabla\!=\!\nabla(\Gamma)$
and $F_{\nu\sigma}\!=\!\tilde\nabla_\nu\w_\sigma-
\tilde\nabla_\sigma\w_\nu\!=\!\partial_\nu
\w_\sigma-\partial_\sigma\w_\nu$.
With $R_{\alpha\mu\nu\sigma}=g_{\alpha\lambda}
R^\lambda_{\,\,\,\mu\nu\sigma}$ then from (\ref{A6})
\bea\label{ss1}
\tilde R_{\alpha\mu\nu\sigma}
&=&
R_{\alpha\mu\nu\sigma}
+\alpha' \Big\{g_{\alpha\sigma} \nabla_\nu\w_\mu- g_{\alpha\nu} \nabla_\sigma \w_\mu
-g_{\mu\sigma}\nabla_\nu \w_\alpha +g_{\mu\nu} \nabla_\sigma\w_\alpha
+g_{\alpha\mu} F_{\nu\sigma}\Big\}
\nonumber\\
&+&\alpha^{' 2} \Big\{
\w^2 (g_{\alpha\sigma} g_{\mu\nu}-g_{\alpha\nu} g_{\mu\sigma} )
+\w_\alpha \,(\w_\nu g_{\sigma\mu}-\w_\sigma g_{\mu\nu})
+\w_\mu (\w_\sigma g_{\alpha\nu} -\w_\nu g_{\alpha\sigma})\Big\}\quad
\eea

\medskip
Since in the Riemannian  geometry  $R_{\alpha\mu\nu\sigma}=R_{\nu\sigma\alpha\mu}$, then from the last equation
\bea\label{tild}
\tilde R_{\alpha\mu\nu\sigma}-\tilde R_{\nu\sigma\alpha\mu}=
\alpha^\prime \Big\{
g_{\alpha\sigma} F_{\nu\mu}+
g_{\mu\nu} F_{\sigma\alpha}+
g_{\alpha\mu} F_{\nu\sigma}
-g_{\alpha\nu} F_{\sigma\mu}
-g_{\mu\sigma} F_{\nu\alpha}
-g_{\nu\sigma} F_{\alpha\mu}\Big\}.
\eea

\medskip
Since $\tilde\Gamma$ is invariant under  (\ref{WGS}) then $\tilde R^\mu_{\,\,\,\nu\rho\sigma}$ is invariant, too, 
and so is the Ricci tensor $\tilde R_{\mu\sigma}\equiv \tilde R^\lambda_{\,\,\,\mu\lambda\sigma}$
of  Weyl geometry. Then from (\ref{A6}), in $d$ dimensions
\bea\label{zz1}
\tilde R_{\mu\sigma}=
R_{\mu\sigma} 
+\alpha'\, 
\Big[\frac{d}{2} F_{\mu\sigma}-(d-2)\nabla_{(\mu} \omega_{\sigma)}
 - g_{\mu\sigma} \nabla_\lambda\omega^\lambda\Big]
+\alpha^{\prime 2}  (d-2) (\omega_\mu\omega_\sigma -g_{\mu\sigma} \omega_\lambda\omega^\lambda)
\eea

\medskip\noindent
with $R_{\mu\nu}$ the Ricci tensor in Riemannian geometry.
Note that $\tilde R_{\mu\nu}-\tilde R_{\nu\mu}=\alpha^\prime d F_{\mu\nu}$.

Then the  Weyl scalar curvature  $\tilde R$ is
\medskip
\bea\label{Rs}
\tilde R=g^{\mu\sigma}\tilde R_{\mu\sigma}=R-2 (d-1)\, \alpha^\prime \, \nabla_\mu \omega^\mu 
-(d-1) (d-2) \,\alpha^{\prime 2} \omega_\mu \omega^\mu.
\eea
in terms of Riemannian scalar curvature $R$.
The  Weyl tensor in Weyl geometry associated to $\tilde R_{\mu\nu\rho\sigma}$,
 is then (with $\tilde C^\mu_{\,\,\,\nu\mu\sigma}=0$)
\bea\label{tc}
\tilde C_{\alpha\mu\nu\sigma}
&=&\tilde R_{\alpha\mu\nu\sigma}
+\frac{1}{d-2} \,\big( g_{\alpha\sigma} \tilde R_{\mu\nu}
 +g_{\mu\nu} \tilde R_{\alpha\sigma}-g_{\alpha\nu}\tilde R_{\mu\sigma}
-g_{\mu\sigma} \tilde R_{\alpha\nu}\big)
\nonumber\\[5pt]
&+&\frac{1}{(d-1)(d-2)}\, \big( g_{\alpha\nu} \,g_{\mu\sigma} - g_{\mu\nu} g_{\alpha\sigma}\big)
\tilde R.
\eea
Replacing  (\ref{ss1})  in the last equation, one finds after some algebra
\bea\label{cct}
\tilde C_{\alpha\mu\nu\sigma}=C_{\alpha\mu\nu\sigma}
+\frac{\alpha'}{d-2} \big(F_{\mu\nu} g_{\alpha\sigma}+F_{\alpha\sigma} g_{\mu\nu}
-F_{\mu\sigma} g_{\alpha\nu}-F_{\alpha\nu} g_{\mu\sigma}\big)
+\alpha'\,g_{\alpha\mu}\, F_{\nu\sigma}.
\eea

\medskip\noindent
From this,  one can show that\footnote{
The square of  Weyl tensor of Riemannian geometry 
used in rhs of (\ref{square}) can be written as (e.g. \cite{Buch})
\be\label{sqr}
C_{\mu\nu\rho\sigma}^2=R_{\mu\nu\rho\sigma}^2-\frac{4}{d-2} R_{\mu\nu}^2
+\frac{2}{(d-1)(d-2)} R^2.\ee}
\bea\label{square}
\tilde C_{\alpha\mu\nu\sigma}^2=
C_{\alpha\mu\nu\sigma}^2+ (d^2-2 d+4)/(d-2) \,\, \alpha^{\prime \, 2} F_{\mu\nu}^2
\eea
 so operators  $\tilde C_{\alpha\mu\nu\sigma}^2$ and $F_{\mu\nu}^2$ are not independent.

Finally, one can show that the structure below corresponds to the
Euler-Gauss-Bonnet term  in Weyl geometry \cite{Tann} (eq.C1) originally found in \cite{Bach}
\medskip
\bea\label{tE}
\tilde G
=\tilde R_{\mu\nu\rho\sigma} \tilde R^{\mu\nu\rho\sigma}-
4\tilde R_{\alpha\beta} \tilde R^{\alpha\beta} +\tilde R^2
+F_{\mu\nu} F^{\mu\nu} \alpha^{\prime \,2} (2 d^2 -7d +8).
\eea

\medskip\noindent 
$\tilde G$ contains  $F_{\mu\nu}^2$, similar to  $\tilde C_{\alpha\mu\nu\sigma}^2$.
For $d=4$, $q=1$, $\alpha^\prime=1/2$ of \cite{Tann}
 one finds the coefficient of $F_{\mu\nu}^2$ is 3,  in agreement with eq.(C1) in \cite{Tann}.

As mentioned in the text, $\tilde \nabla$ does not keep manifest Weyl covariance 
when acting on a curvature tensor. To implement this, denote 
\medskip
\bea
\hat\nabla_\mu g_{\alpha\beta}\equiv(\tilde\nabla_\mu+ \alpha q\,\,\w_\mu) g_{\alpha\beta}
\eea 
where $q$ is the charge of $g_{\alpha\beta}$, hence replace $\partial_\mu g_{\alpha\beta}\ra 
(\partial_\mu +\alpha q \omega_\mu) g_{\alpha\beta}$ in the action of
$\tilde\nabla$, so
\bea\label{hh}
\hat \nabla_\mu g_{\alpha\beta}=0.
\eea
i.e. the theory is metric  with respect to  the new differential operator $\hat\nabla$.
Under (\ref{WGS}) one finds  a Weyl-covariant transformation:
$\hat \nabla_\mu' g_{\alpha\beta}'=\Sigma^q \hat\nabla_\mu g_{\alpha\beta}$.
More generally, for any tensor $T$ of charge $q_T$,  with $T^\prime=\Sigma^{q_T} T$ (the indices of $T$ are
 not shown)
we have
\bea
\hat \nabla_\mu T\equiv (\tilde \nabla_\mu + \alpha \,q_T\,\w_\mu)T\quad\Ra\quad
\hat\nabla'_\mu T^\prime=\Sigma^{q_T} \hat \nabla_\mu T.
\eea
 $\tilde\nabla$ has the usual geometric action (with $\tilde\Gamma$).  Similar Weyl covariant
transformation  applies to  $\hat\nabla_\mu\hat\nabla^\mu T$, etc.
Also note that $\hat F_{\mu\nu}=\hat\nabla_\mu\w_\nu-\hat\nabla_\nu\w_\mu=
\partial_\mu\w_\nu-\partial_\nu\w_\mu=F_{\mu\nu}$.

At a more fundamental level, in the  basis $e_a=e_a^\mu \partial_\mu$ where
$e_a^\mu e_b^\nu \eta^{ab}=g^{\mu\nu}$, with  $\eta_{ab}$ the Minkowski metric,
one has that
$\hat \Gamma_{ab}^c e_c=\hat \nabla_a e_b=(\tilde\nabla_a-\alpha q/2\, \w_a) e_b$ 
 because $e_b=e_b^\mu \partial_\mu$ has Weyl charge $-q/2$
i.e. half of that of $g^{\mu\nu}=e_a^\mu e_b^\nu\eta_{ab}$ ($\partial_\mu$ and $dx^\mu$ have zero charge).
 Here we denoted $\w_a=\w_\mu\,e^\mu_a$.
Then  the Riemann tensor in Weyl geometry in the "basis" with a "hat"  (called the
natural Weyl "basis") is (with notation  $\alpha'\equiv \alpha q/2$):
\medskip
\bea
\hat R^a_{\,\,\,bcd} \, e_a=[\hat\nabla_c, \hat\nabla_d] \,e_b
=[\tilde\nabla_c,\tilde\nabla_d]\, e_b - \alpha^\prime F_{cd} e_b=
\tilde R^a_{\,\,\,bcd}\,e_a - \alpha^\prime\delta_b^a\, F_{cd}\, e_a
\eea
This gives
\bea\label{hattilde}
\hat R^\rho_{\,\,\,\mu\nu\sigma}=\tilde R^\rho_{\,\,\,\mu\nu\sigma}-\alpha^\prime \delta_\mu^\rho \, F_{\nu\sigma}
\eea
%
which "removes" the last term in the first line of  eq.(\ref{A6}). Using 
$\hat R_{\alpha\mu\nu\sigma}=g_{\alpha\rho} \hat R^\rho_{\,\,\,\mu\nu\sigma}$ then
\medskip
\bea\label{hhh}
\hat R_{\alpha\mu\nu\sigma}=\tilde R_{\alpha\mu\nu\sigma}-\alpha^\prime g_{\alpha\mu} \,F_{\nu\sigma},
\qquad
\hat R_{\mu\sigma} = \tilde R_{\mu\sigma}-\alpha^\prime F_{\mu\sigma},
\qquad
\hat R=\tilde R.
\eea

\medskip\noindent
We can write these in terms of 
Riemannian geometry counterparts,\,with (\ref{ss1}), (\ref{zz1}),(\ref{Rs})
\bea\label{ss1p}
\hat R_{\alpha\mu\nu\sigma}
\!\!\!\!&=&\!\!\!
R_{\alpha\mu\nu\sigma}
+\alpha' \Big\{g_{\alpha\sigma} \nabla_\nu\w_\mu- g_{\alpha\nu} \nabla_\sigma \w_\mu
-g_{\mu\sigma}\nabla_\nu \w_\alpha +g_{\mu\nu} \nabla_\sigma\w_\alpha
\Big\}
\nonumber\\[5pt]
&+&\!\!\!\alpha^{' 2} \Big\{
\w^2 (g_{\alpha\sigma} g_{\mu\nu}-g_{\alpha\nu} g_{\mu\sigma} )
+\w_\alpha \,(\w_\nu g_{\sigma\mu}-\w_\sigma g_{\mu\nu})
+\w_\mu (\w_\sigma g_{\alpha\nu} -\w_\nu g_{\alpha\sigma})\Big\},\quad
\nonumber\\[5pt]
\hat R_{\mu\sigma}&=&\!\!\!\!
R_{\mu\sigma} 
+\alpha'\, 
\Big[\frac{d-2}{2} F_{\mu\sigma}-(d-2)\nabla_{(\mu} \omega_{\sigma)}
 - g_{\mu\sigma} \nabla_\lambda\omega^\lambda\Big]
+\alpha^{\prime 2}  (d-2) (\omega_\mu\omega_\sigma -g_{\mu\sigma} \omega_\lambda\omega^\lambda),
\nonumber\\[5pt]
\hat R&=& R-2 (d-1)\, \alpha^\prime \, \nabla_\mu \omega^\mu 
-(d-1) (d-2) \,\alpha^{\prime 2} \omega_\mu \omega^\mu.
\label{hhhp}
\eea

\medskip\noindent
Further,  $\hat R_{\mu\sigma}-\hat R_{\sigma\mu}=\alpha^\prime (d-2) F_{\mu\sigma}$
and 
\bea
\hat R_{\alpha\mu\nu\sigma}-\hat R_{\nu\sigma\alpha\mu}=
\alpha^\prime \Big\{
g_{\alpha\sigma} F_{\nu\mu}+
g_{\mu\nu} F_{\sigma\alpha}
-g_{\alpha\nu} F_{\sigma\mu}
-g_{\mu\sigma} F_{\nu\alpha}
\Big\}.
\eea

\medskip
From  (\ref{hhh}),  (\ref{tc})  and a similar  version of eq.(\ref{tc})  but
 in the "hat" notation one finds
\bea
\tilde C_{\mu\nu\rho\sigma}=\hat C_{\mu\nu\rho\sigma}
+\frac{\alpha'}{d-2} \big( g_{\mu\sigma} F_{\nu\rho}+g_{\nu\rho} F_{\mu\sigma} 
-g_{\mu\rho} F_{\nu\sigma} -g_{\nu\sigma} F_{\mu\rho}\big)+\alpha' g_{\mu\nu} \,F_{\rho\sigma}.
\eea

\medskip\noindent
Comparing this last equation  to (\ref{cct}) one finds that
\bea\label{CCC}
\hat C_{\alpha\nu\rho\sigma}=C_{\alpha\nu\rho\sigma}.
\eea

\medskip\noindent
Therefore, the Weyl tensor in Weyl geometry (in the "hat"/Weyl natural basis) is identical to the
Riemannian Weyl tensor and so is independent of $F_{\mu\nu}$, unlike $\tilde C_{\alpha\mu\nu\sigma}$ 
of (\ref{cct}), (\ref{square}).

The Euler-Gauss-Bonnet term
of Weyl geometry in the "hat" notation is then (note the position of summation indices)
\medskip
\bea\label{a19}
\hat G=\hat R_{\mu\nu\rho\sigma} \hat R^{\rho\sigma\mu\nu}-
4 \hat R_{\mu\nu} \hat R^{\nu\mu} +\hat R^2.
\eea

\medskip\noindent
Like the Weyl tensor above,  $\hat G$ does not contain $F_{\mu\nu}^2$ anymore 
 (unlike in eq.(\ref{tE})). $\hat G$ is found from  $\tilde G$ of (\ref{tE})
after some algebra, by  using (\ref{hattilde}) and (\ref{hhh}).
Note that this expression of
$\hat G$ is similar to its  Riemannian geometry counterpart recovered for 
 $\w_\mu=0$, since then $\hat R_{\mu\nu\rho\sigma}\ra R_{\mu\nu\rho\sigma}$,
$\hat R_{\mu\nu}\ra R_{\mu\nu}$, $\hat R\ra R$,  etc.

Further, by direct calculation of the rhs of the equation below (in terms of
their Riemannian counterparts, see
eqs.(\ref{hhhp}) and (\ref{sqr})) one can show that
\smallskip
\bea\label{ccc1}
\hat C_{\mu\nu\rho\sigma}\hat C^{\mu\nu\rho\sigma}=
\hat R_{\mu\nu\rho\sigma}\,\hat R^{\rho\sigma\mu\nu}-\frac{4}{d-2}\,\hat R_{\mu\nu}\hat R^{\nu\mu}
+\frac{2}{(d-1)(d-2)}\hat R^2.
\eea

\medskip\noindent
This relation is a generalisation to Weyl geometry of the similar relation in Riemannian case,
see eq.(\ref{sqr}).
Finally, using (\ref{a19}) in the last equation, we find
\medskip
\bea\label{CCC1}
\hat C_{\mu\nu\rho\sigma}\hat C^{\mu\nu\rho\sigma}=
\hat G+\frac{d-3}{d-2} \,\Big[4 \hat R_{\mu\nu} \,\hat R^{\nu\mu}-\frac{d}{d-1} \hat R^2\Big].
\eea

\medskip\noindent
Like (\ref{a19}), (\ref{ccc1}) from which it was derived,
eq.(\ref{CCC1}) also extends to Weyl geometry a similar relation of Riemannian geometry.

\bigskip
\counterwithin{equation}{subsection}
\setcounter{equation}{0}
\def\theequation{B-\arabic{equation}}
\def\thesubsection{B}
\def\thefigure{B-\arabic{figure}}
\def\thelabel{B}

 \subsection{ Weyl gauge symmetry current}\label{B}

$\bullet$  For an arbitrary Weyl gauge invariant action we show there is a
non-trivial,  conserved  current $J^\mu$ in $d=4$, information used in Section~\ref{3.2}.
Consider  a Weyl gauge transformation in $d=4$ dimensions:
\bea
 g_{\mu\nu}'=\Sigma^q\,g_{\mu\nu},\qquad
\tilde\phi'=\Sigma^{q_\phi}\tilde \phi,\quad
 q_{\tilde\phi}=-\frac{q}{2}; \quad
\omega_\mu'=\omega_\mu-\frac{1}{\alpha} \partial_\mu \ln\Sigma
\eea
where $\tilde\phi$ is here some scalar field.
For an infinitesimal transformation $\delta \Sigma$ 
\medskip
\bea\label{var}
\delta  g_{\mu\nu}'=\delta(\ln\Sigma^q)\,g_{\mu\nu}',\quad
\delta\tilde\phi'=-\frac12\,\delta(\ln\Sigma^{q})\,\tilde\phi',\quad
\delta\omega_\mu'
=-\frac{1}{\alpha}\partial_\mu\, \delta\ln\Sigma.
\eea

\medskip
Consider a Weyl gauge invariant {\it total} action given by the sum $W_g+W$, where
$W_g$ is the Weyl gauge field kinetic term while $W$
 is the remaining  action that can depend
on $\omega_\mu$ but not on $\hat F_{\mu\nu}$, hence
\be
W_g=-\frac14\int d^4 x \sqrt{g}\, \hat F_{\mu\nu}^2 
\qquad
W=\int d^4 x \sqrt{g} \,L
\ee
$W_g$ and $W$ are each Weyl gauge invariant.
Under (\ref{var}) 
\bea
\delta W=\int d^4 x \sqrt{ g'}\,
\Big[
-\frac{1}{2}\, T^{\mu\nu}\,\delta  g_{\mu\nu}' + J^\mu\,\delta \omega_\mu' +
\frac{1}{\sqrt{g}}\frac{\delta W}{\delta \tilde\phi'} \,\delta\tilde\phi'\Big],
\eea

\medskip\noindent
where $T^{\mu\nu}$ and $J^\mu$ are the energy-momentum tensor associated with $W$ and
the Weyl gauge symmetry  current, respectively.
The last term in $\delta W$ vanishes by the equation of motion for $\tilde\phi$. 
Since $W$ is Weyl gauge invariant ($\delta W=0$) and using (\ref{var}),
then  (after removing the "prime'' notation):
\smallskip
\bea
0=\delta W\!\!\!&\!\!\equiv\!\! &\!\!\!\!
\int\!\! \sqrt{g}\Big[
-\frac{1}{2} T^{\mu\nu} g_{\mu\nu}
\,\delta \ln\Sigma^q -\frac{1}{\alpha}J^\mu\,\partial_\mu \delta\ln\Sigma\Big]
\!=\!\!\!\int\!\!\sqrt{g} \Big[-\frac12\,T^\mu_\mu
+\frac{1}{\alpha q} \nabla_\mu J^\mu\Big]\delta\ln\Sigma^q.\quad\quad
\eea

\medskip\noindent
where we used that in Riemannian geometry
$\sqrt{g}\nabla_\mu J^\mu=\partial_\mu (J^\mu \sqrt{g})$.
Therefore, for a Weyl gauge invariant action
\bea\label{T}
T_\mu^\mu=\frac{2}{\alpha q}\nabla_\mu J^\mu.
\eea

\medskip\noindent
used in the text, eq.(\ref{traceT}).
Finally, from the total action $W+W_g$ one can easily write the equation of motion for $\omega_\mu$:
\medskip
\bea
J^\mu + \nabla_\sigma F^{\sigma\mu}=0
\eea

\medskip\noindent
with Riemannian $\nabla_\sigma$.
Multiply  this equation by $\sqrt{g}$ and  apply $\partial_\mu$
and  use the antisymmetry of $F^{\sigma\mu}$ to find  $\nabla_\mu J^\mu=0$, i.e.
there is  a conserved current onshell.

\vspace{0.3cm}
\bigskip\noindent
$\bullet$
Let us now take a  particular case for the Weyl action in $d=4$ (no matter):
\bea\label{SSS}
W &\equiv& 
\int d^4x \sqrt{g} \Big\{
\frac{1}{4! \,\xi^2} \hat R^2-\frac{1}{\eta^2} \hat C_{\mu\nu\rho\sigma}^2\Big\}
\\[5pt]
&=&
\int  d^4x  \sqrt{g}\, \Big\{
\frac{-1}{12\xi^2}
\,\phi^2\,\Big[ R-3\,q\,\alpha \nabla_\mu\omega^\mu
-\frac32\,q^2\,\alpha^2\,\omega_\mu\omega^\mu\Big]
-\frac{\phi^4}{4!\,\xi^2}
-\frac{1}{\eta^2} C_{\mu\nu\rho\sigma}^2\Big\},
\eea

\medskip\noindent
where we linearised $\hat R^2$ as explained in Section~\ref{3.4}
 with a scalar (dilaton) $\phi$ of equation of motion $\phi^2=-\hat R$.
The Euler-Gauss-Bonnet $\hat G$ term was not added to the above action
 since it does not change the equations of motion here.
In the second line we used a Riemannian notation (with $\nabla_\mu$ given by the Levi-Civita connection)
and  the relation between $\hat R$, $\hat C_{\mu\nu\rho\sigma}$
and their Riemannian counterparts (without a hat), see Appendix~\ref{A},
eqs.(\ref{hhhp}), (\ref{CCC}).
We find a current 
\medskip\bea\label{jjj}
J_\mu=\frac{1}{\sqrt{g}} \frac{\delta W}{\delta \omega^{\mu}}=
-\frac{\alpha\,q}{4\xi^2}\,(\partial_\mu-\alpha\,q \omega_\mu)\phi^2.
\eea

\medskip\noindent
The total action  $W+W_g$ gives 
the following equation of motion for $\omega_\mu$ 
\bea\label{eom}
\sqrt{g}\,\Big\{\,
\frac{\alpha^2\,q^2}{4\,\xi^2}\phi^2 \, \omega^\rho -\frac{\alpha\,q}{4\xi^2}\,\nabla^\rho\phi^2
+\nabla_\sigma F^{\sigma\rho}\Big\}=0,
\eea

\medskip\noindent
This equation is Weyl gauge invariant (expected, since  the action  is invariant).
Apply $\partial_\rho$ on the last equation, use 
$\sqrt{g}\,\nabla_\sigma F^{\sigma\mu}= \partial_\sigma\, (\sqrt{g}\,F^{\sigma\mu})$
and the antisymmetry of $F^{\sigma\rho}$, to find
\bea\label{div}
\nabla_\mu J^\mu=0.
\eea
Therefore, there exists a non-trivial, conserved current onshell.

Let us now check eq.(\ref{T}) for our case. 
From action (\ref{SSS}) one has 
\medskip
\be\label{tr}
g^{\mu\nu}\frac{\delta W}{\delta g^{\mu\nu}}=
\frac{\sqrt{g}}{12\,\xi^2}
\,\Big[ 
- 3 \Box\phi^2 + 3 \alpha q \nabla^\rho (\omega_\rho\phi^2)
\Big] 
\ee

\medskip\noindent
where we used that $\phi^2=-\hat R$. Here $\Box=\nabla_\mu \nabla^\mu$ is in Riemannian 
notation.
This result is actually valid  for the total action $W+W_g$
since
the contribution  to the trace by the (conformal) gauge kinetic term $F_{\mu\nu}^2 \sqrt{g}$
is vanishing. Therefore
\bea\label{opi}
T_\mu^\mu=\frac{2}{\sqrt{g}} g^{\mu\nu} \frac{\delta W}{\delta g^{\mu\nu}}=
\frac{-1}{2\,\xi^2} \,\nabla^\rho( \nabla_\rho - \alpha q\omega_\rho)\phi^2.
\eea
We thus have
\bea
T_\mu^\mu=\Big(\frac{2}{\alpha q}\Big)\, \nabla_\mu J^\mu,
\eea

\medskip\noindent
in agreement with general result (\ref{T}).

\bigskip
\noindent
{\bf Acknowledgements:\,\,\,}

\noindent
The author thanks Luca Ciambelli (Perimeter Institute),  Cezar Condeescu  (IFIN Bucharest),
Christopher T. Hill (Fermilab),
Weizhen Jia (Illionis University, Urbana), Andrei Micu (IFIN Bucharest), Mikhail Shaposhnikov (University
of Lausanne) for  discussions on Weyl conformal geometry.
This work was supported by a grant of  Ministry
of Education and Research (Romania), CNCS-UEFISCDI, project number
PN-III-P4-ID-PCE-2020-2255.

\end{document}